**Simultaneous Digital Communication and Deformation Sensing over a Single Stretchable Interconnect**

*Yuji Isano and Hiroki Ota**

Yuji Isano, Hiroki Ota
Department of Mechanical Engineering, Yokohama National University, Yokohama, Kanagawa, Japan

Hiroki Ota
Graduate School of System Integration, Yokohama National University, Yokohama, Kanagawa, Japan
Institute of Advanced Sciences, Yokohama National University, Yokohama, Kanagawa, Japan

Funding: JSPS KAKENHI Grant-in-Aid for Transformative Research Areas (No. 24H00890)
Grant-in-Aid for Scientific Research A (24H00296)
Grant-in-Aid for JSPS Fellows (23KJ0991)

Abstract. Stretchable hybrid electronics integrate rigid solid-state electronics with stretchable materials and structures to achieve both high deformability and stable electronic performance. However, most existing systems treat stretchability only as a mechanical attribute without exploiting device deformation to encode its own mechanical state. This problem arises from adapting conventional rigid circuit architectures to stretchable substrates, affording a loss in compatibility with the sensors required for strain measurement. This study addresses this issue by proposing a communication-integrated deformation sensing architecture for stretchable hybrid devices. In the proposed approach, standard universal asynchronous receiver-transmitter digital signals transmitted between rigid nodes are amplitude-modulated by strain-induced resistance changes in stretchable liquid metal interconnects. By reading both amplitude changes and digital patterns, the system enables simultaneous digital communication and self-deformation sensing without requiring additional stretchable sensing elements. The architecture is demonstrated in a multi-node system and applied to wearable sensing and self-deformation mapping devices. By extending the integration of rigid circuits and soft elements from the hardware level to the system level, this study provides a novel design paradigm for stretchable electronic systems that inherently utilize their own deformation as functional information.

# 1. Introduction

Stretchable devices are crucial for applications such as wearable electronics, soft robotics, stretchable displays, and bio-sensing interfaces.[1] Among these, stretchable hybrid devices combine rigid electronic components with flexible or stretchable substrates and wiring, enabling the stable signal processing, digital communication, and advanced computational capabilities of conventional rigid electronic components while accommodating large mechanical deformation.[2] Their core advantage lies in integrating the mature functionality and reliability of rigid electronics with the mechanical compliance of stretchable systems.

The development of conventional stretchable hybrid devices has primarily focused on using stretchability to enhance the freedom of attachment sites and improve conformability. Consequently, numerous stretchable counterparts of conventional electronic systems have been demonstrated, including physical sensor arrays,[3-6] light-emitting diode (LED) arrays,[7,8] electromyography (EMG) sensors,[9] photoplethysmography (PPG) sensors,[10] and microcontroller unit (MCU) platforms,[11,12] which maintain stable operation even under deformation. Meanwhile, the miniaturization of rigid electronics now enables rigid devices to achieve these advantages as well. As such, stretchable hybrid devices are gradually losing their unique value.

One unique advantage of stretchable devices is their ability to deform in unison with an object and utilize that deformation as information. For example, such information can represent the device's shape, posture, or the motion of an object they are in close contact with. Capturing this information is fundamentally difficult for rigid devices that do not deform; it is a function that can only be realized by stretchable devices. Despite this, stretchable hybrid devices capable of simultaneously performing their primary electronic functions while converting their own deformation into information, e.g., by measuring their state of deformation, have rarely been developed.

The inability to utilize deformation as functional information stems from architectural constraints specific to stretchable hybrid devices, which result from directly adapting the design principles of rigid circuits. Rigid circuits achieve high packaging density via three-dimensional mounting, which uses fine wiring and stacked multiple layers. By contrast, stretchable devices face challenges in miniaturizing elastic conductors and forming reliable vias and interlayer connections within elastomeric substrates. Although attempts have been made to create multilayer stretchable structures, their packaging density remains orders of magnitude lower than that of rigid circuits. Consequently, these devices inevitably rely on two-dimensional packaging, which effectively spreads the components out on a flat surface. Even simply incorporating the same functionality as the rigid device consumes a significant portion of the limited packaging area. Under these constraints, such elastomeric wiring serves merely as a flexible component of the rigid circuit, whereas its intrinsic electromechanical response remains largely unchanged.

Under these circumstances, deformation measurement requires dedicated deformable sensing elements. However, this approach presents two fundamental challenges. First, the additional components and wiring further reduce the available mounting area. Second, the increased number of components and loose contacts, which can cause failures or signal degradation, reduces the overall reliability of the system and negates the stability that rigid circuits typically provide. This loss of reliability is independent of the mounting area and cannot be resolved by simply adding more flexible components or layers. Therefore, a circuit and communication architecture optimized for stretchable devices is required—one that incorporates deformation measurement by using the intrinsic responses of soft materials instead of treating them merely as stretchable wiring, while maintaining continuity with rigid-circuit technology.

In this study, we develop a communication-integrated strain measurement architecture for stretchable hybrid devices (Figure 1). This architecture addresses the aforementioned area and wiring constraints without adding new elastic elements for measurement but by utilizing the stretchable wire for inter-rigid circuit communication also as a strain sensor. In this scheme, the intrinsic resistance–strain response of the stretchable conductor itself is functionalized as the modulation mechanism of digital communication; the low-hysteresis and cyclically stable liquid metal wiring ensures accurate and repeatable operation. Additionally, a layered rigid–soft hybrid structure, which relieves the strain concentration around rigid nodes via local stiffness control, is introduced as the physical platform that maintains this architecture under large deformation. This approach establishes a new circuit and communication paradigm for stretchable hybrid devices, expanding such devices from mere carriers of rigid electronic functions that withstand deformation to electronic systems that can utilize their own expansion and contraction as functional information.

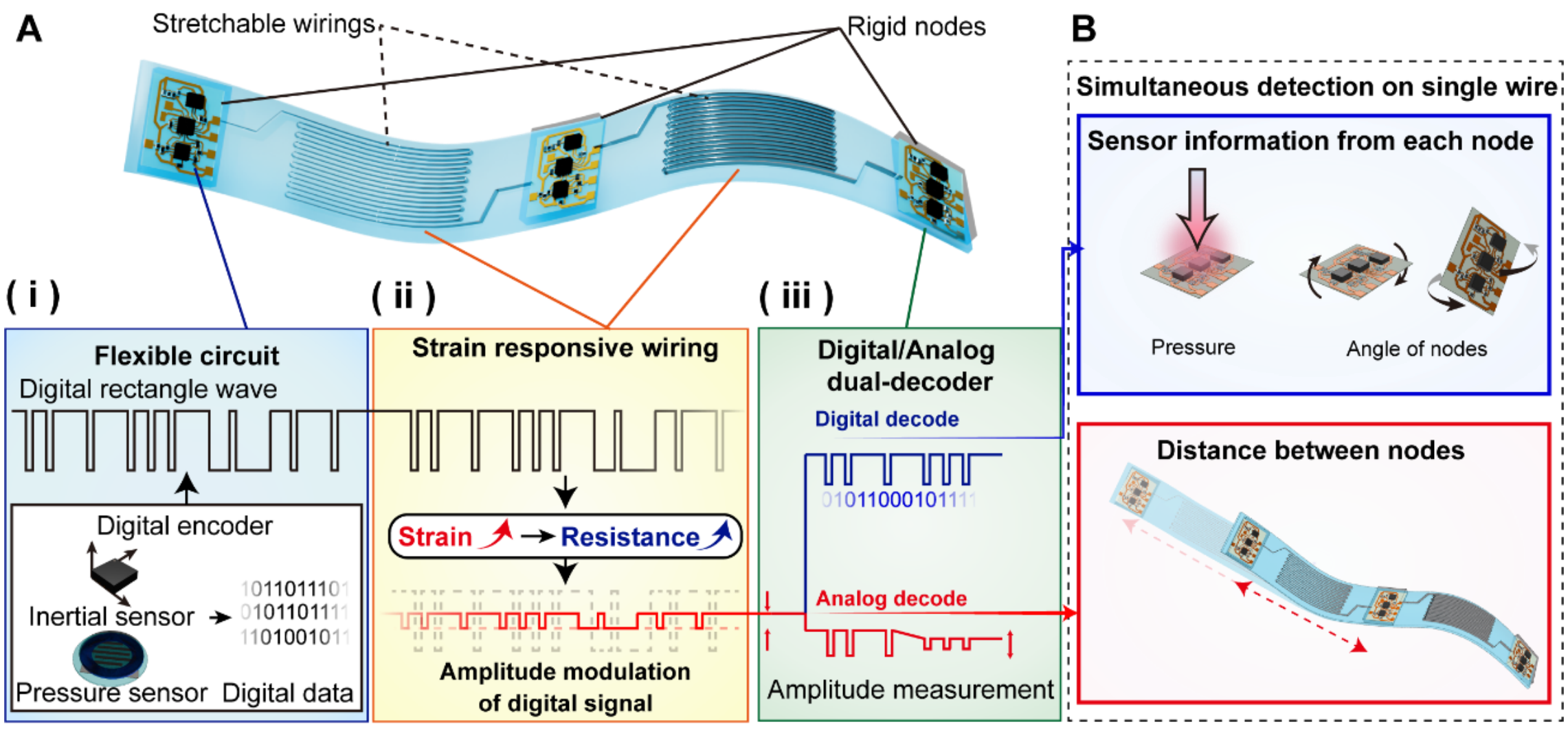

**Figure 1.** Architecture for simultaneous communication and deformation measurement in stretchable hybrid devices. A. System overview: (i) Signals acquired by the flexible circuit (rigid node) on the device are encoded into digital packets and output via universal asynchronous receiver-transmitter (UART) communication. (ii) Subsequently, the output signal is amplitude-modulated based on the changes in the resistance of the stretchable wiring. (iii) Finally, the amplitude of the modulated digital signal and the digital data are read by the decoder circuit at the far end. This three-stage system enables both the digital transmission of measurement data and the measurement of the inter-node distance variations caused by the changes in wiring resistance using the same wiring (shown in Figure 1B).

## 2. Results and Discussion

### 2.1. Architecture of the Proposed Multimodal Communication System

The proposed communication system that simultaneously detects digital signals and deformation of the communication line comprises three elements: a transmitter circuit on a stretchable hybrid device, a communication line whose resistance varies with deformation, and a receiver-side multi-decoder (**Figure 2A**). These components are mounted on a single flexible printed circuit board (PCB) and bonded to the stretchable substrate. The transmitter circuit encodes data such as sensor measurements according to a specially designed packet structure and transmits it using an existing universal asynchronous receiver-transmitter (UART) communication system. In this process, the signal takes the form of a digital waveform, where logics 1 and 0 are represented by the supply voltage (VCC) and GND, respectively. The multi-decoder that receives the signal comprises four units: (1) an inverting attenuator, (2) an analog measurement circuit, (3) a comparator circuit, and (4) a digital readout circuit. The multi-decoder processed the signal according to the following procedure: (1) The transmitted digital data was first input into the inverting attenuator via an elastic wire whose resistance varies owing to strain. The digital signal, which had an amplitude ranging from GND to the source voltage (VCC), was attenuated by the inverting attenuator according to Equation 1, in proportion to the ratio of the feedback resistance to the resistance of the elastic wire, and converted into a signal with an analog amplitude centered at half the VCC (Figure 2B).

$$V_{out} = V_{ref} - \frac{R_{ref}}{R_{wire}}(V_{in} - V_{ref}) \quad (1)$$

(2) The resistance of the stretchable wiring was calculated by reading this signal with an analog-to-digital converter (ADC) in the analog measurement circuit and comparing it to the VCC to determine the amplitude attenuation rate and estimate the wiring strain. (3) The analog-modulated signal was fed into a comparator circuit after its amplitude was measured. The comparator demodulated the analog-modulated signal into a digital signal with an amplitude ranging from GND to VCC (Figure 2C). (4) Finally, the demodulated digital signal was read by the UART communication unit of the receiving microcontroller. The transmitted data was obtained by decoding the data in the digital signal according to its packet structure. The system kept pace with signal transitions at the commonly used communication rate of 115,200 bps (bits per second) and was used to successfully perform analog amplitude modulation and digital demodulation in response to strain (Figure 2B and 2C).

Digital communication using the proposed multi-decoder circuit enables strain detection and helps to alleviate the constraints on the wiring materials in stretchable devices. The inverting attenuator enables the input impedance to be adjusted by changing the feedback resistance; hence, stretchable materials with a wide range of resistances can be used for device wiring. Figure 2D shows the measured output amplitude of the digital signal (VCC = 3.3 V) as a function of the combination of the reference resistance ($R_{ref}$) and wiring resistance of the multi-decoder. The results show a consistent signal amplitude matching the circuit simulation results (**Figure S1**) accounting for the internal resistance of the integrated circuit (IC) across the entire measured resistance range, up to $R_{ref}$ = 100 kΩ and $R_{wire}$ = 1.5 MΩ. In addition, an experiment was conducted in which packets were transmitted continuously at 115,200 bps every 10 ms for 25 s. All transmitted data passed through the multi-decoder and was received without losses. These results indicate that the proposed strain-digital multiplex communication architecture using a multi-decoder can accommodate a wide range of stretchable conductive materials as wiring, including not only low-resistance materials such as liquid metals and highly conductive composites, [13, 14] but also conductive polymers with resistances in the several-kΩ range [15, 16] and carbon-based composites with resistances of several hundred kΩ.

Considering these characteristics, this study adopted liquid metal (oxidized galinstan) as the wiring material, prioritizing its excellent measurement stability as a strain sensor over its digital communication characteristics. The liquid metal meandering wiring used in this study exhibits negligible hysteresis in the strain–resistance response during elongation and relaxation and maintains stable resistance changes even after 200 cycles of deformation (**Figure S2** and **S3**). Such stability of individual elements is crucial in system designs that combine numerous elements because it affects the accuracy of the resistance–strain conversion when the sensor is integrated into a system.

Wiring strain was measured using a multi-decoder with $R_{ref}$ set at 30 Ω to match the resistance of the liquid metal wiring. The method for measuring the amplitude is shown in **Figure S4**. Voltage readings were taken at regular intervals from an analog amplitude-modulated signal, and the data was classified into HIGH and LOW by comparing the values with the threshold of VCC/2. Subsequently, the current HIGH and LOW values were calculated by averaging the results of the previous 10 measurements, and the amplitude was calculated by subtraction. Figure 2E shows the changes in the wiring strain and modulated wave amplitude at different communication speeds. Consequently, strain measurement with a sensitivity of approximately −6.7 mV/strain% (at VCC = 3.3 V), unaffected by the communication speed, was achieved for transmission speeds ranging from 4,800 bps to 115,200 bps. The theoretical minimum measurable value for the 12-bit ADC used in this study is 0.8 mV at a supply voltage of 3.3 V, and 3 mV for a 10-bit ADC. These values are lower than the amplitude change corresponding to 1% strain; hence, a resolution of 10 bits or higher is sufficient for strain measurement. Figure 2F shows the results of measuring the response speed of the strain measurement. The amplitude measurement results were smoothed using a 1 s moving average. Although the moving average processing caused a slight delay in the change in resistance relative to the deformation, amplitude fluctuations that followed a deformation rate of 200%/s could be observed. This indicates that the proposed method is useful for applications that frequently involve large deformations, such as hand motion recognition and device deformation measurement.

Using this method enables the detection of deformation in sections not directly connected to the master readout unit, as well as the reading of element data, with minimal error. **Figure S5** shows an overview of serial communication using relay transmission. Figure 2G shows the results of simultaneously measuring the digital data from two units and the strain between them via this relay communication using three units. In this measurement, each unit was equipped with a digital-to-analog converter (DAC), which outputs arbitrary voltages instead of a sensor, and that output was measured and transmitted as digital data. The strains between the terminal and intermediate units [Figure 2G(i)] and between the intermediate and master units [Figure 2G(ii)] were varied from 0 to 50%, respectively. For each strain condition, the DACs on each unit were varied across five distinct levels, and their outputs were transmitted as digital data [Figure 2G(iii) and (iv)]. Consequently, the two wiring strains were detected independently without any crosstalk. Furthermore, the process by which the DACs on the two units sequentially changed their voltage outputs was read without any loss or distortion, realizing the simultaneous transmission of strain and data via three-unit serial relay communication. Conversely, because time-based random sampling may sometimes capture HIGH signals that have not fully risen, relatively large noise is superimposed on the strain signals (**Figure S6**, **Supplemental Note 1**). We believe this can be resolved by improving the sampling method.

Although various stretchable hybrid devices have been developed in the past, many of them have focused on the physical design or functionality of the devices. However, studies focusing on unique communication architectures for these devices are limited. In terms of communication multiplexing, analog sensor data have been multiplexed using alternating current (AC) and frequency-division multiplexing.[17, 18] In terms of new communication protocols for deformable devices, research has been conducted on pulse-digital communication modeled after neural signals [19] and communication based on flexible circuits that can be disconnected and reconnected;[20, 21] however, simultaneous digital communication and deformation measurement over the same wiring on a stretchable device have not yet been achieved. Existing methods either require analog modulation and frequency resolution for communication multiplexing [17, 18] or rely on dedicated protocols or specialized flexible structures;[19–21] thus, they either (i) deviate from the standard digital communication framework or (ii) require the addition of dedicated elements or materials for strain measurement. By contrast, in this study, communication and deformation measurement are integrated onto a single wire without using additional stretchable elements while maintaining standard UART digital communication.

Furthermore, existing systems require specific materials and device configurations on the flexible device side, as well as complex measurement equipment for signal encoding and decoding. By contrast, the system developed in this study offers several practical advantages. First, by constructing the core of the system using highly reliable rigid circuits and existing digital communication protocols, we achieved easy reproducibility and broad applicability across various systems. Second, our method of digitally converting sensor outputs at each node enables the use of any sensor readable by an MCU; furthermore, an inverting attenuation buffer circuit enables the integration of strain-responsive wiring with a wide resistance range, from 50 Ω to 1 MΩ, into the system. Third, as no additional flexible elements are required for strain measurement, the limited mounting area of stretchable hybrid devices can be effectively utilized. In these respects, the proposed system can be used as a highly versatile communication method with performance comparable to a general-purpose communication protocol for stretchable devices.

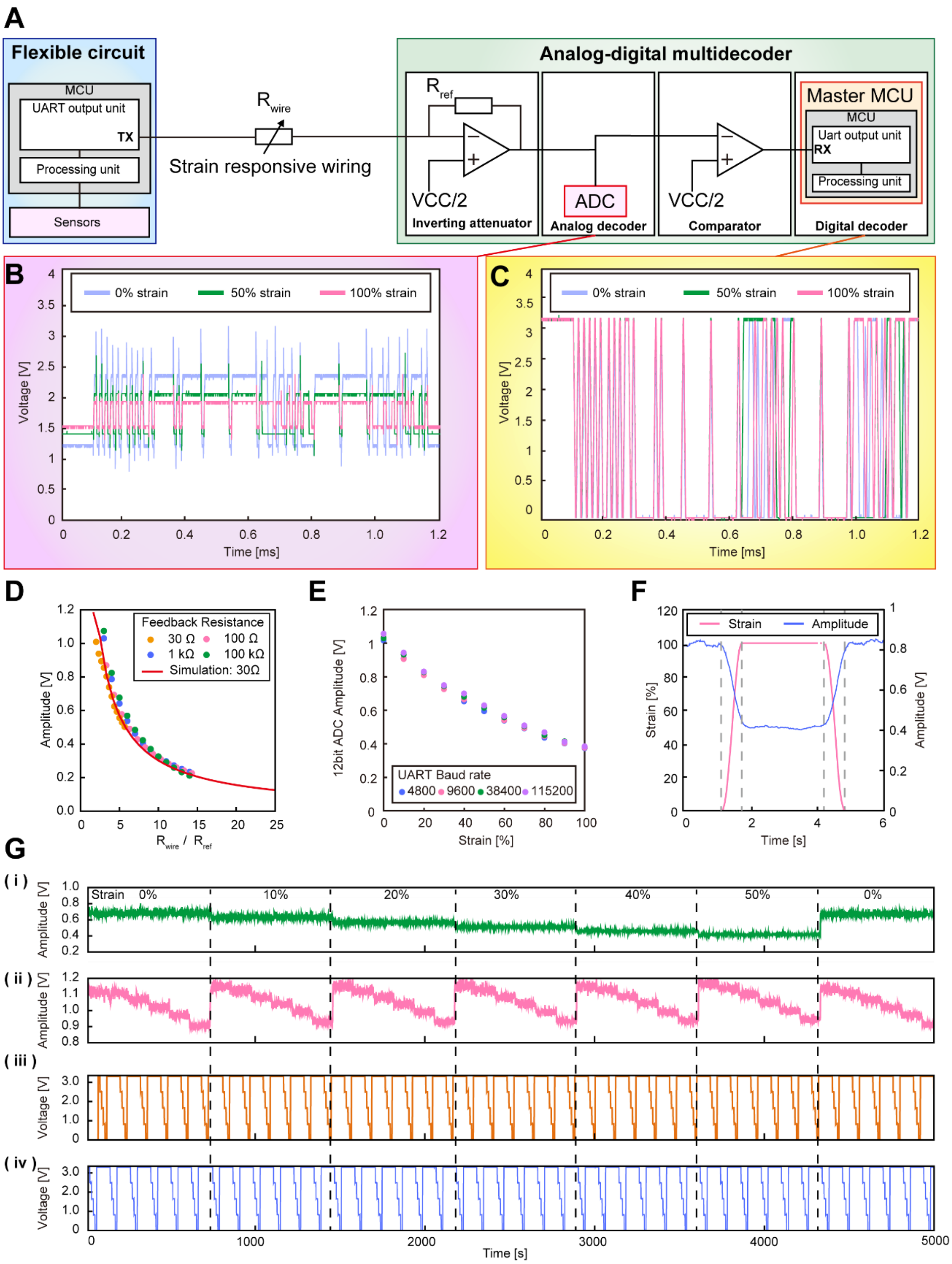

**Figure 2.** Characteristics of the simultaneous communication and strain measurement architecture. A. Schematic of the communication method. The sensor output measured on the flexible circuit was converted into a digital packet in UART format by the MCU and output as a square wave signal with an amplitude ranging from VCC to 0 V. The receiver-side multi-decoder comprises an inverting attenuator, an ADC, a comparator, and an MCU, which perform amplitude modulation, amplitude reading, digital demodulation, and digital reading, respectively. B, C. Signals immediately before the analog decoder (B) and digital decoder (C). Owing to the strain-

responsive wiring and the inverting attenuator, amplitude attenuation corresponding to strain occurs immediately before the analog decoder. The comparator decodes the signal into a digital waveform with the amplitude of the VCC (3.3 V) to 0 V immediately before the digital decoder. D. Measurement of the amplitude response for multiple $R_{ref}$ values using fixed resistors instead of strain-responsive wiring. All results were similar to simulations that accounted for the internal resistance of the IC. E. Relationship between the strain and analog amplitude at different communication speeds using liquid metal wiring. In the 4,800–115,200 bps range, the amplitude attenuation was nearly linear with respect to the strain. F. Time variation of strain and the measured amplitude in liquid metal wiring. The amplitude was smoothed using a 1 s moving average. G. Digitally received measurement results of the analog signal amplitude and the DAC output simulating a sensor in a device comprising three nodes and wiring connected in series. (i) Upstream amplitude, (ii) downstream amplitude, (iii) upstream DAC output, (iv) downstream DAC output. All signals were detected without crosstalk.

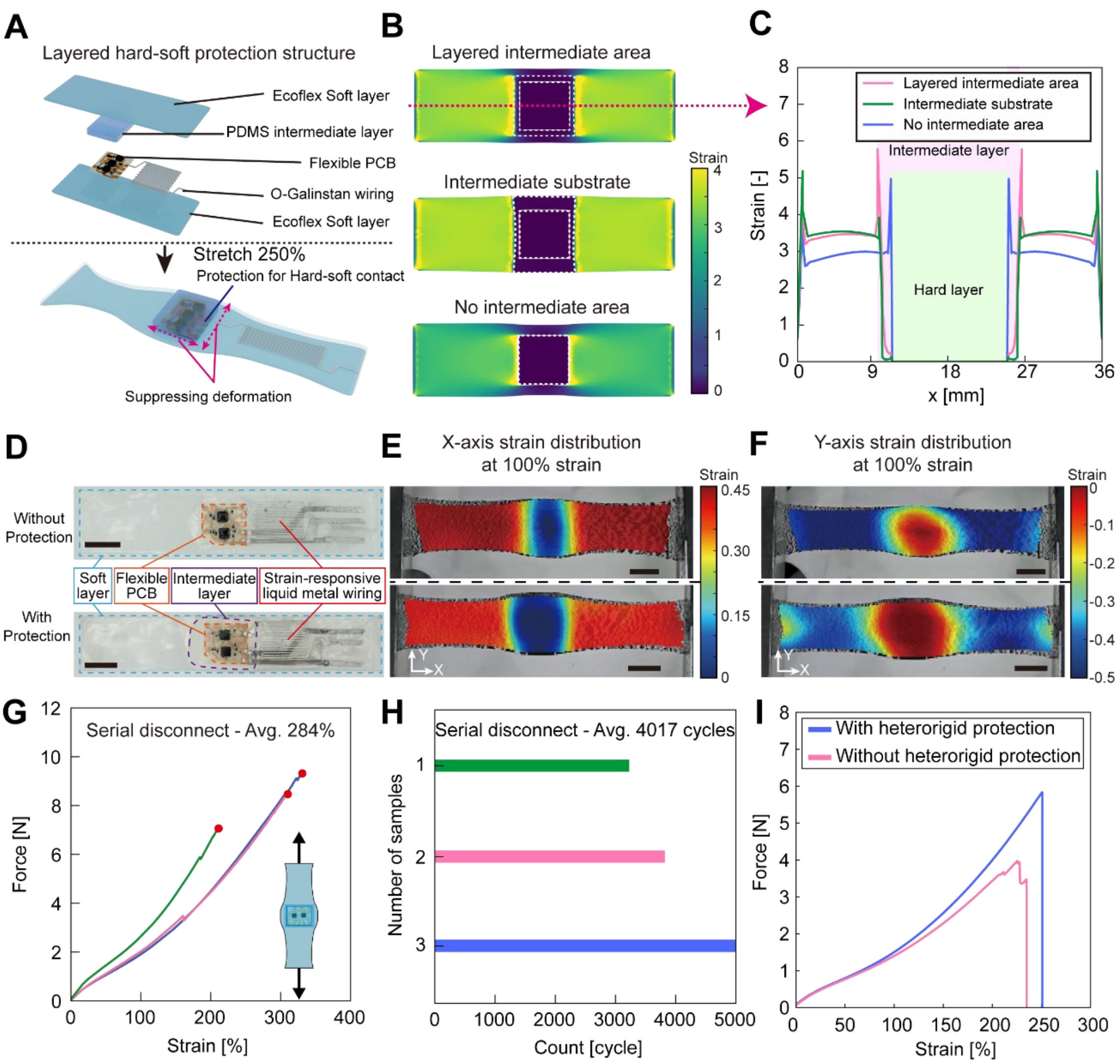

Figure 3. Rigid node protection using a layered rigid–soft hybrid structure. A. Overview of the layered rigid–soft hybrid structure. An intermediate layer is formed by laminating a PDMS layer, which has a stiffness intermediate between those of the rigid node and the soft substrate. B, C. Strain distributions at 100% elongation in the layered structure, the substrate-embedded

intermediate-layer structure, and the structure without an intermediate layer, as determined by finite element analysis (FEA) simulation: two-dimensional (B) and along the centerline (C). The layered structure exhibited a strain distribution at the hard–intermediate layer interface that was nearly identical to that of the substrate with an intermediate layer. D. Evaluation device for a layered hard–soft hybrid structure. The device outputs a digital signal that increases as the structure deforms. E, F. Visualization of the actual strain distribution at 100% elongation using digital image correlation (DIC): X-axis direction (E) and Y-axis direction (F). The strain-relieving effect of the intermediate layer can be observed in the horizontal and vertical directions of the tensile axis. Scale bar: 10 mm. G. Results of deforming the device until communication was lost. The device maintained communication up to an average strain of 284% (n = 3). H. Results of repeatedly stretching the device to 100% until communication was lost. The device maintained communication for an average of 4,017 cycles (n = 3). I. Results of a tensile test conducted until the device physically broke.

### 2.2. Layered Rigid–Soft Hybrid Assembly for Integrating High-Performance Circuits under Deformation

The proposed architecture is designed to operate under large deformation conditions; hence, implementation techniques to protect rigid nodes from the high strains being measured are essential. Distributing high-functionality ICs across a stretchable substrate enables the realization of a multimodal system for communication and deformation detection, as shown in Figure 2. In the surrounding areas of these rigid elements, rapid strain gradients occur as the device deforms, which can easily result in wire breaks.[22,23] Therefore, in this study, we introduced a layered hard–soft hybrid structure that controls local stiffness by drop-casting a stiffer silicone resin around an IC-mounted flexible PCB attached onto a softer silicone substrate. Compared with substrate fabrication methods that integrate materials with different stiffnesses, this approach enables a simple fabrication process while maintaining the effective protection of the circuit (**Figure 3A**).

First, a finite element analysis (FEA) simulation was conducted to clarify the strain concentration mitigation performance around the hard layer in a laminated hard–soft hybrid structure. The simulations were conducted for three configurations: a layered intermediate structure, a substrate-embedded intermediate area, and no intermediate area (**Figure S7**). The results showed that in the layered rigid–soft hybrid structure, the strain concentration at 100% elongation was greater at the intermediate–soft layer interface than that in the substrate-embedded structure. At the flexible substrate–interlayer interface, strain is suppressed to nearly zero, similar to the substrate-embedded structure. By contrast, strain concentrates at the flexible substrate–soft layer interface in structures without a protective layer (Figure 3B and C). Based on these results, the layered rigid–soft hybrid structure is expected to provide protection for hard components equivalent to that of conventional substrate-embedded protective structures.

Next, the actual strain distribution of the device, with and without a protective layer, was analyzed using digital image correlation (DIC). A device containing a single unit was stretched to 100% of its original length to compare the differences in strain distribution between devices with and without a protective layer (Figure 3D–F). Regarding the strain distribution in the X-axis direction, the device without a protective layer exhibited greater strain near the hard layer compared with that with a protective layer (Figure 3E). Furthermore, the Y-direction compressive strain gradient near the hard layer was significantly larger in the device without a protective layer, indicating that the protective layer shields the periphery of the flexible PCB from compressive deformation (Figure 3F). Suppressing the tensile strain in the X direction prevents separation of the wiring from the rigid layer, whereas suppressing the compressive strain in the Y direction prevents bending deformation of the flexible PCB, thereby preventing the generation of delamination forces associated with differences in the bending stiffness of the layered structure.

This protective structure contributes to maintaining the multimodal communication system under deformation. The devices with a protective layer maintain their communication and measurement functionality up to an average elongation of 284% across the three measured samples (Figure 3G, **Figure S8**). By contrast, those without a protective layer lost their communicative function at an average elongation of 173% (**Figure S9**). Furthermore, when the devices were subjected to 5,000 cycles of 100% elongation, the protected devices withstood an average of 4,017 cycles (Figure 3H, **Figure S10**), whereas the devices without a protective layer failed after an average of 888 cycles (**Figure S11**). Upon applying deformation until the device physically broke, the device without a protective layer exhibited multiple sudden force fluctuations in the stage immediately preceding fracture. This indicates that a change in the overall stiffness of the system occurred owing to the delamination of the hard layer when the unprotected device fractured. Meanwhile, no such force fluctuations were observed in the protected device. This suggests that no delamination of the hard layer occurred until just before fracture and that the device fracture resulted from the failure of the soft layer. Notably, one device with a protective layer exhibited a significantly smaller maximum deformation, and its deformation force characteristics were similar to those of the device without a protective layer. This may be owing to individual variations in the hard layer or the mounting conditions during the tensile test, which may have caused the concentration of tensile stress near the hard layer, resulting in a stress concentration similar to that observed in the device without a protective layer.

Thus, this layered rigid–soft hybrid protective structure has been shown to significantly improve the deformation tolerance and cyclic durability of devices equipped with multimodal communication systems. Various methods have been proposed to improve stretchability and reduce stress concentration. For example, a device achieved up to 700% elongation via structural engineering of the rigid layer, [24] and a substrate achieved 200% elongation by continuously changing the stiffness.[25] However, these existing methods require prior processing of the shape and materials of the substrate, as well as dedicated, optimized fabrication systems such as three-dimensional (3D) printers. Although the method described in this study is inferior to these devices in terms of stretchability, it achieved an elongation resistance exceeding 250% (average 284%, n = 3) through simple pick-and-place bonding followed by the post-processing application of a protective resin. This method simplifies the fabrication of rigid–flex structures and contributes to the integration of rigid ICs and flexible materials.

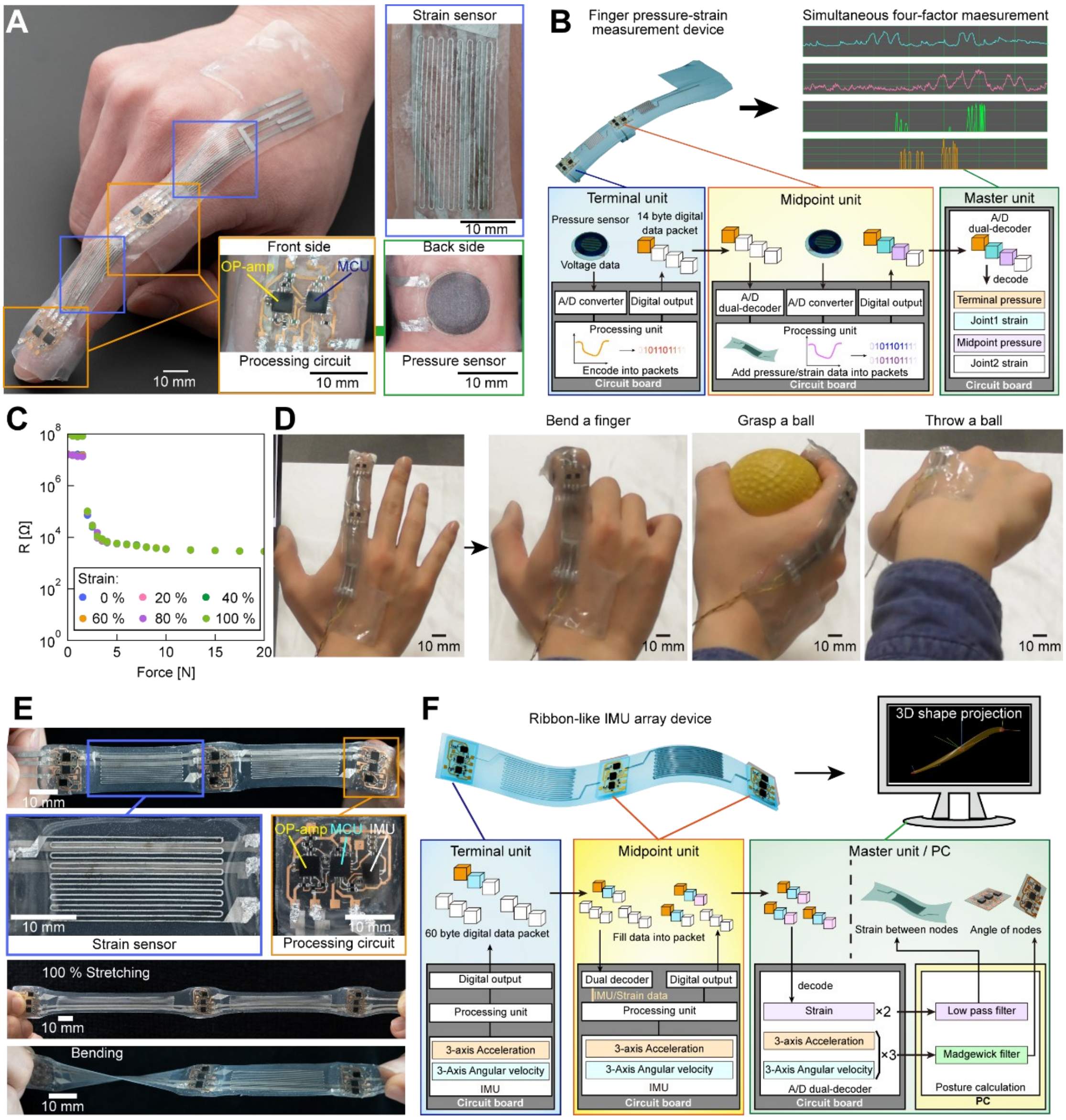

Figure 4. Demonstration of a multimodal communication architecture for simultaneous strain measurement. A. Device for simultaneous measurement of finger pressure and bending. Rigid nodes and pressure sensors are mounted at the fingertip and midway along the finger to measure and transmit finger pressure, whereas simultaneous strain measurement is conducted to detect finger joint bending using a single wire. B. System overview. The unit at the fingertip only transmits digital data. The mid-section unit receives digital data and simultaneously performs strain detection; it then digitizes the pressure value it measured, the strain between the tip node and itself, and the data received from the tip node and transmits this information to the master. C. Relationship between the strain state of the multimodal communication wiring and the output values of the pressure sensors. Through digital conversion, the sensors exhibited a constant output regardless of the strain on the wiring. D. A demonstration of the ball-gripping experiment. E, F. A self-shape measurement device using acceleration sensors. Using a simultaneous communication and strain measurement architecture with 6-axis inertial sensors mounted on each of the three nodes, the device measures the orientation of each node and the distance between nodes and projects the data in 3D.

## 2.3. Integration of the Proposed Multimodal Communication System in Practical Applications

We demonstrated the usefulness of our multimodal communication systems in stretchable device applications using two types of practical devices: a stretchable device that simultaneously measures finger pressure and finger joint flexion and a system that performs 3D mapping of the angles of each unit and the distances between units. Stretchable sensor systems for the human hand are in high demand owing to their high degree of freedom. In addition, given the benefits of measuring various physical elements such as shape, motion, and palm surface pressure, various measurement systems using stretchable devices have been developed. However, incorporating multiple types of sensors requires different measurement systems. In addition, the mounting area of the device is constrained by the finger width. Consequently, integrating different measurement systems simultaneously is difficult. Therefore, we integrated the communication lines and finger deformation sensors by utilizing the proposed system and relay communication, thereby achieving the simultaneous measurement of finger pressure and bending using a single-system architecture. **Figure 4A** shows the actual device, and Figure 4B shows a schematic diagram of the system. The device comprises two communication units connected in series; the communication lines are used to measure the flexion of the proximal interphalangeal (PIP) and metacarpophalangeal (MP) joints. The measurements from the resistive force sensors mounted on each unit were acquired via a voltage divider circuit and an ADC on the unit. The measurements were then converted to digital data according to a specified packet structure and transmitted through liquid metal wiring. The force sensor incorporated into the device was fabricated using a deformable membrane structure in which a conductive film and electrodes were bonded by spacers (**Figure S12**). The sensor exhibited a decrease in resistance in response to the applied force and, despite the presence of hysteresis, demonstrated high sensitivity in the range 2.5–10 N (**Figure S13**). This sensor was connected to a multimodal measurement system using a single unit, and pressure measurements were performed with the communication line under deformation. Consequently, a stable output was obtained in the high sensitivity range of 2.5 N or higher, regardless of the deformation state (Figure 4C).

Next, the communication protocol for the multimodal measurement and communication system designed to maintain data accuracy in practical applications is described. The data packet structure is shown in Figure S14. A packet comprises a magic number indicating the start of the packet, ID of the node that last edited the data (source ID), total number of nodes, data, and a cyclic redundancy check (CRC). The total data length of a packet depends on the amount of data; it was 14 bytes for simultaneous measurement of finger deformation and pressure (**Figure S14**A) and 60 bytes for deformation mapping (Figure S14B). The transmitted data was read one byte at a time using a sliding window of a predetermined data length. The beginning of the packet was identified by the presence of the magic number. Subsequently, the system verified that the source ID and total number of nodes complied with the specifications, and the integrity of the entire dataset was finally confirmed using the CRC (**Figure S15**). These multiple data integrity checks ensured that no corrupted data resulting from relay communication or analog modulation and demodulation was introduced into the system.

Herein we demonstrated the simultaneous measurement of finger force and flexion using the newly developed system to track various hand movements. The multimodal communication system detected the changes in finger force resulting from finger flexion and various ways of gripping a ball, as well as larger movements such as throwing a ball, without any delay (**Figure S16, Supplemental Video 1**).

Next, the self-shape mapping system is described. In self-shape recognition for stretchable devices, a key advantage of stretchable hybrid devices is the ability to measure the angle of each unit by placing high-performance inertial sensors on them. However, it is necessary to recognize the relative positional relationships between nodes to achieve shape recognition; this was previously measured using strain sensors or external cameras operating on a separate system from the inertial sensors. In this study, three-unit relay communication via a multimodal measurement and communication system was employed to determine the tilt and relative distance of the units using a single-system approach. Figure 4E shows the actual device. The device operated precisely even under 100% deformation and transmitted data on the tilt of each unit and the changes in the distance between them. Figure 4F shows an overview of the system. Each unit was equipped with an inertial sensor in addition to the communication system, enabling the measurement of six-axis motion consisting of three-axis acceleration and three-axis angular velocity. The data were transmitted to the master unit according to a predetermined packet structure. The tilt of each unit and the distance between units were calculated from the data transmitted from the master unit to the PC and projected using 3D mapping software (**Figure S17, Supplemental Video 2**). From these demonstrations, the operational capability and high versatility of the multimodal measurement and communication system on an actual device were shown.

Although numerous devices capable of hand shape and fingertip pressure recognition [26, 27] and self-deformation mapping [6, 28, 29] have been reported, these processes either require devices and sensors specifically designed for such applications or employ structures that directly mount conventional rigid circuit architectures. By contrast, the proposed system offers broad versatility. First, because measurement values are digitally converted within the unit, the system can be operated with any type of analog-to-digital sensor—not only resistive sensors, as demonstrated in this study, but also capacitive [30,31] and voltage-output [32,33] sensors. Second, because it uses flexible circuit units capable of accommodating high-performance components, it can result in the development of highly noise-resistant, stretchable devices that can incorporate sensors with low output and high noise sensitivity—such as electrochemical sensors[34]—without the need for long-distance wiring. Third, processing the data enables the identification of units whose data have been edited. This suggests the potential to expand beyond the serial connections of units to various wiring patterns. Fourth, since the system uses general-purpose MCUs and digital communication systems, it can, in principle, be applied to other existing digital communication protocols, such as the Inter-Integrated Circuit (I2C) and Serial Peripheral Interface (SPI), in any system that can perform digital communication (in this study, it was demonstrated using UART).

Therefore, existing rigid electronics resources can be seamlessly repurposed for stretchable device systems using this architecture. This study presents a new direction for extending the concept of integrating stretchable devices and rigid electronics, which has thus far been limited to the hardware level, to the software and systems level.

However, this method has several limitations. First, because strain measurement relies on resistance changes, the measured signal is affected by temperature variations and aging drift, necessitating calibration and compensation for high-precision absolute measurements. Second, the strain signal is superimposed with noise resulting from time-based random sampling (Supplemental Note 1). This noise could potentially be reduced by synchronizing the sampling with the communication process. Third, the demonstration in this study is limited to circuits comprising bulk devices and serial connections of up to three nodes. These are not fundamental limitations of this architecture and can be improved through implementation optimization.

## 3. Conclusions

In this study, we proposed a circuit and communication architecture for stretchable hybrid devices that utilizes the deformation of communication lines as information. Distance changes between rigid elements were detected by amplitude-modulating the digital signal voltage used for communication between them in response to changes in the wiring resistance. The digital signal was then demodulated and received via an existing digital communication interface. Consequently, digital and analog signals were both detected without crosstalk, thereby simultaneously enabling strain detection with a sensitivity of −6.7 mV/strain% using a single wire and the reception of digital signals. Furthermore, owing to the layered rigid–soft hybrid structure, the device maintained its communication functionality up to an average elongation of 284% (n = 3).
The study findings represent a significant step toward expanding the functionality of stretchable hybrid devices that integrate existing solid-state electronics with soft devices. The core of the proposed architecture comprises a combination of existing rigid circuits and communication protocols, offering high reproducibility, versatility, and compatibility with rigid circuits. Furthermore, by utilizing a single stretchable interconnect for multiple purposes, the challenge of limited mounting area in stretchable hybrid devices is mitigated, facilitating higher device density and the integration of additional flexible circuit elements. The measurement circuit in this study was constructed using bulk devices to demonstrate the proof of concept; however, future studies will develop smaller devices with higher functional integration by introducing circuit and communication architectures optimized for stretchable hybrid devices, for example, starting from the IC design stage.

## 4. Methods

Development and Evaluation of a Multimodal Strain Measurement and Communication System: During the evaluation phase of the communication system, the transmitter circuit, liquid metal wiring, and receiver multi-decoder circuit were not integrated but connected using commercially available metal wires. The transmitter flexible circuit and relay circuit utilized an operational amplifier (ADA4528, Analog Devices) in addition to the 8-bit ADC, UART communication circuit, and comparator circuit built into a microcontroller (ATtiny1616, Microchip Technology). The multi-decoder circuit was constructed using an operational amplifier (LT1678, Analog Devices), a comparator (NJU7109F, Nisshinbo Micro Devices), a UART receiver circuit, and a 12-bit ADC mounted on a microcontroller board (Raspberry Pi Pico, Raspberry Pi Foundation). The actual communication waveform modulated by the strain-sensing wiring and then demodulated by the digital decoder was recorded using a digital oscilloscope (PicoScope 2205A, Pico Technology). The relationship between resistance values and amplitude changes for different feedback resistors in the inverting amplifier circuit was verified using a circuit built on a breadboard. Device deformation was investigated by stretching the device using a single-axis stage (OSMS26-50, Sigma). Electrical circuit simulations were performed using LTSpice (Analog Devices).

*Device Fabrication*: The device was fabricated using the method shown in Figure S18. A two-component silicone resin (Ecoflex 00-50, Smooth-on; A:B = 1:1), which serves as the material for the soft substrate, was molded using a mold fabricated with a 3D printer (Form3+, Formlabs) [**Figure S18** (i)]. Before demolding the flexible substrate, a mask was fabricated using a polyimide film (thickness: 50 $\mu$m), which was then used to print liquid metal paste circuits [Figure S18 (ii)]. Next, a 33 wt% Ecoflex-hexane solution was applied via hand spraying to encapsulate the backside wiring [Figure S18 (iii)]. After peeling the flexible substrate from the mold, a polyimide film was laminated onto the substrate [Figure S18(iv)], and a mask for the communication lines capable of strain detection was patterned using a laser marker (MD-1010, Keyence) [Figure S18(v) and (vi)]. Next, excimer treatment was performed to improve the adhesion between the liquid metal and the substrate [Figure S18(vii)]; after applying the liquid metal paste, the polyimide film mask was peeled off to form the wiring pattern [Figure S18(viii) and (ix)]. Subsequently, via holes were formed by carefully cutting only the Ecoflex in the center layer with scissors, thereby exposing the contact points of the backside wiring pattern [Figure S18(x)]. Next, a flexible PCB with the control circuit formed on it was attached [Figure S18(xi)], and the liquid metal wiring was connected to the flexible PCB [Figure S18(xii)]. For encapsulation, PDMS, which serves as a protective layer, was drop-cast onto the flexible PCB and cured to form an intermediate layer [Figure S18(xiii)]. Finally, a 33 wt% Ecoflex-hexane solution was spray-coated using a hand sprayer to encapsulate the device, thereby completing its fabrication [Figure S18(xiv) and (xv)].

Physical Characteristics of the Device: Simulations of the nonlinear strain characteristics of hard–soft hybrid devices were performed using COMSOL Multiphysics. The strain state of the actual device was visualized using the DIC method.[5] The strain–force curves were obtained using a tensile testing machine (EZ-LX, Shimadzu Corp.).

*Simultaneous Measurement of Finger Pressure and Deformation*: The device was fabricated using the method shown in Figure S18. The pressure sensor, fabricated using the method described below, was attached to the device simultaneously with the control circuit. After which, the pressure sensor section was rolled into a cylindrical shape and bonded with an adhesive (Silipoxy, Smooth-on) to create a ring-shaped device. The output from the pressure sensor was detected by a microcontroller mounted on a flexible circuit board using a voltage divider circuit. The device was connected via a wired connection to an external multi-decoder; the acquired data were ultimately processed and visualized using Python.

The flexible pressure sensor was fabricated using the method shown in Figure S12. First, a conductive layer was prepared. Tetrahydrofuran (FUJIFILM Wako) and a styrene–butadiene–styrene copolymer (SBS, Sigma-Aldrich) were mixed in a vial at a mass ratio of THF:SBS = 15:1 and dissolved completely. Next, conductive carbon (SuperP, MTI Japan) was mixed with the solution at a mass ratio of SBS:SuperP = 100:15 and blended using a rotary-orbital mixer (ARE-310, THINKY) [Figure S12A(i)]. This blend was spin-coated onto a 125 μm PET film [Figure S12A(ii)], and the conductive layer was completed by heating it in an oven at 80 °C for 30 min [Figure S12A(iii)]. The separator layer was fabricated by laser-cutting a 125 μm PET film [Figure S12B(i) and (ii)].

The electrode layer was fabricated by printing silver ink onto a 50 μm PET film. First, a polyimide film (7414L, 3M) was laminated onto the PET film and cut with a laser marker to form a mask [Figure S12C(i)]. Electrodes were fabricated by applying conductive silver ink (FC-435, Fujikura Kasei) over this mask and drying it in an oven [Figure S12C(ii) and (iii)]. Subsequently, a resistive pressure sensor was fabricated by sequentially bonding these three layers together.

Deformation Mapping Demonstration: The devices were fabricated using the method described in Figure S18. In addition to the encoder unit, each flexible circuit unit included a 6-axis inertial sensor (ICM-42605, TDK), which transmitted measurements of angle and gravitational acceleration. Based on the acquired inertial data, the Madgwick filter was applied using Python software to estimate the angle of each unit. In addition, the strain in the device was estimated by comparing the acquired strain measurements with previously obtained reference values. The results of the angle and strain estimations were displayed on a screen using a 3D visualization system based on Three.js.

## Conflict of Interest

All of the authors declare that they have no competing interests.

## Declaration of AI Assistance

In this study, ChatGPT (OpenAI) was used to assist in creating control programs for the microcontrollers, as well as the signal processing and visualization programs written in Python and JavaScript.

Acknowledgments

This research is supported by the JSPS KAKENHI Grant-in-Aid for Transformative Research Areas (No. 24H00890), Grant-in-Aid for Scientific Research A (24H00296), and Grant-in-Aid for JSPS Fellows (23KJ0991). The authors thank professor Kazuhide Ueno for providing access to the tensile tester.

Author Contributions
Yuji Isano: Conceptualization, Methodology, Software, Investigation, Data curation, Formal analysis, Writing-Original draft, Visualization, Funding acquisition.
Hiroki Ota: Conceptualization, Supervision, Project administration, Writing-Review & editing, Funding acquisition.

**Data Availability Statement**
The corresponding author will provide the source data for this study upon appropriate request.

Supporting Information

# Simultaneous Digital Communication and Deformation Sensing over a Single Stretchable Interconnect

*Yuji Isano, Hiroki Ota**

## Supplemental Note 1

Effect of packet size on strain sensing noise

Figure S6 shows the distribution of the signal packet size, amplitude, HIGH level, and LOW level with no strain applied to the wiring. These results indicate that the amplitude noise is caused by sampling on the HIGH side. Furthermore, when the signal packet size is small (Figure S6A), the range of the voltage measurement values is narrower than when the packet size is large (Figure S6B).
Based on these results, it can be concluded that the noise observed in strain measurements is attributable to the sampling method. In digital communications, a certain amount of time is required for the voltage to accurately reflect the correct amplitude during high-to-low transitions. However, the simple time-based sampling and threshold-based high–low discrimination used in this study include these inaccurate signals in the measured values. The inclusion of these values during signal switching is presumed to cause the distribution on the high side to become wider. The stability of the LOW-side value is due to the specifications of universal asynchronous receiver-transmitter (UART) communication. The UART communication voltage is fixed at HIGH during non-communication periods; the LOW-side signal is generated by inverting and attenuating this voltage. In the system used in this study, the communication time is sufficiently shorter the non-communication time; therefore, the LOW-side signal is dominated by the voltage during non-communication periods, when no signal fluctuation occurs. Consequently, values from signal transitions are less affected, producing a stable output. This result is supported by differences in the voltage distribution patterns depending on packet size. When the packet size is large, the number of high–low transitions increases, and the probability of sampling occurring at the moment of a signal transition also increases; consequently, there are more opportunities to observe a lower-than-normal potential on the HIGH side. Conversely, owing to the sufficiently long non-communication periods, the LOW side is virtually unaffected by fluctuations caused by packet size.

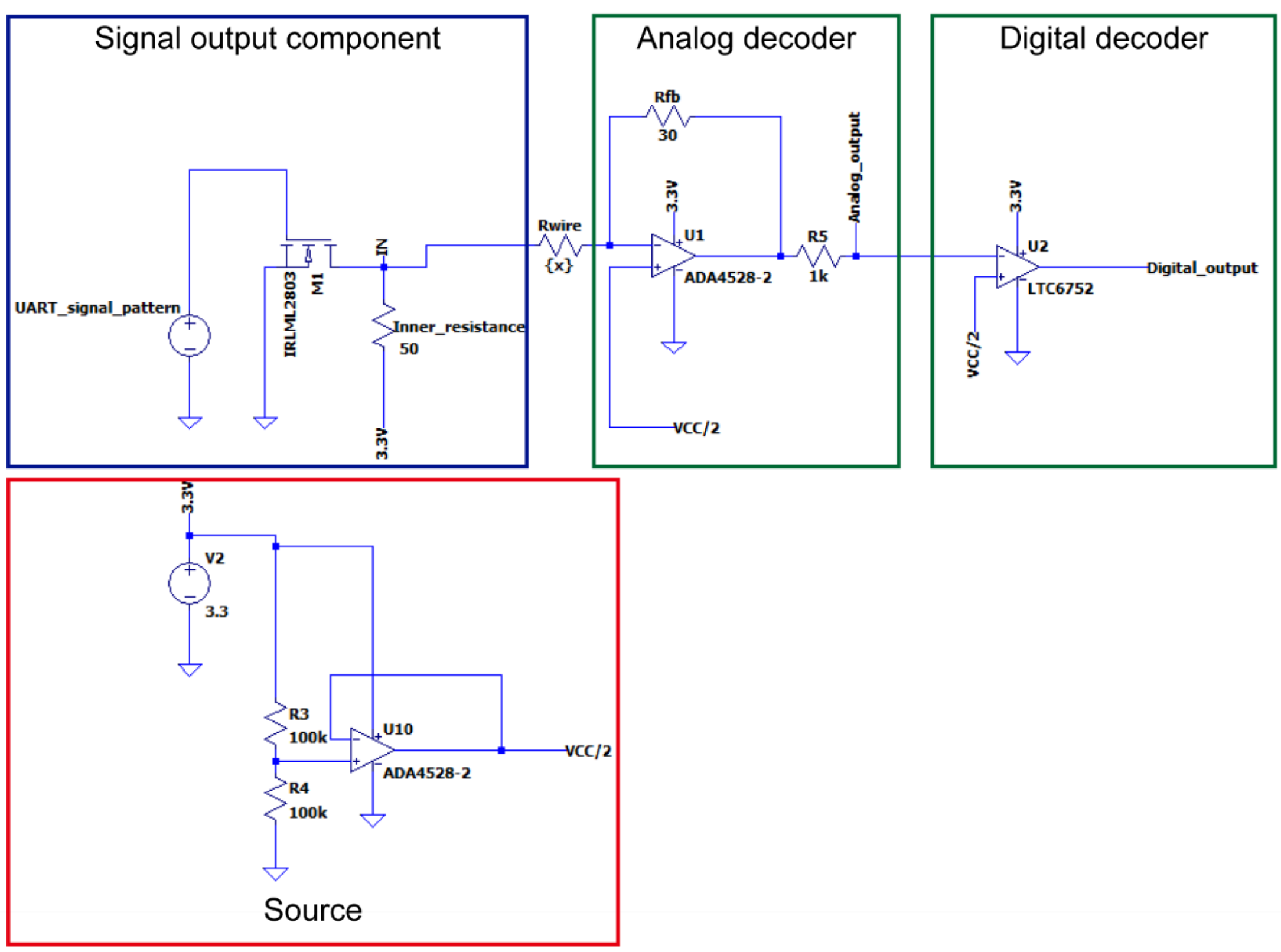

Figure S1. Circuit diagram for simulation using LTSpice.

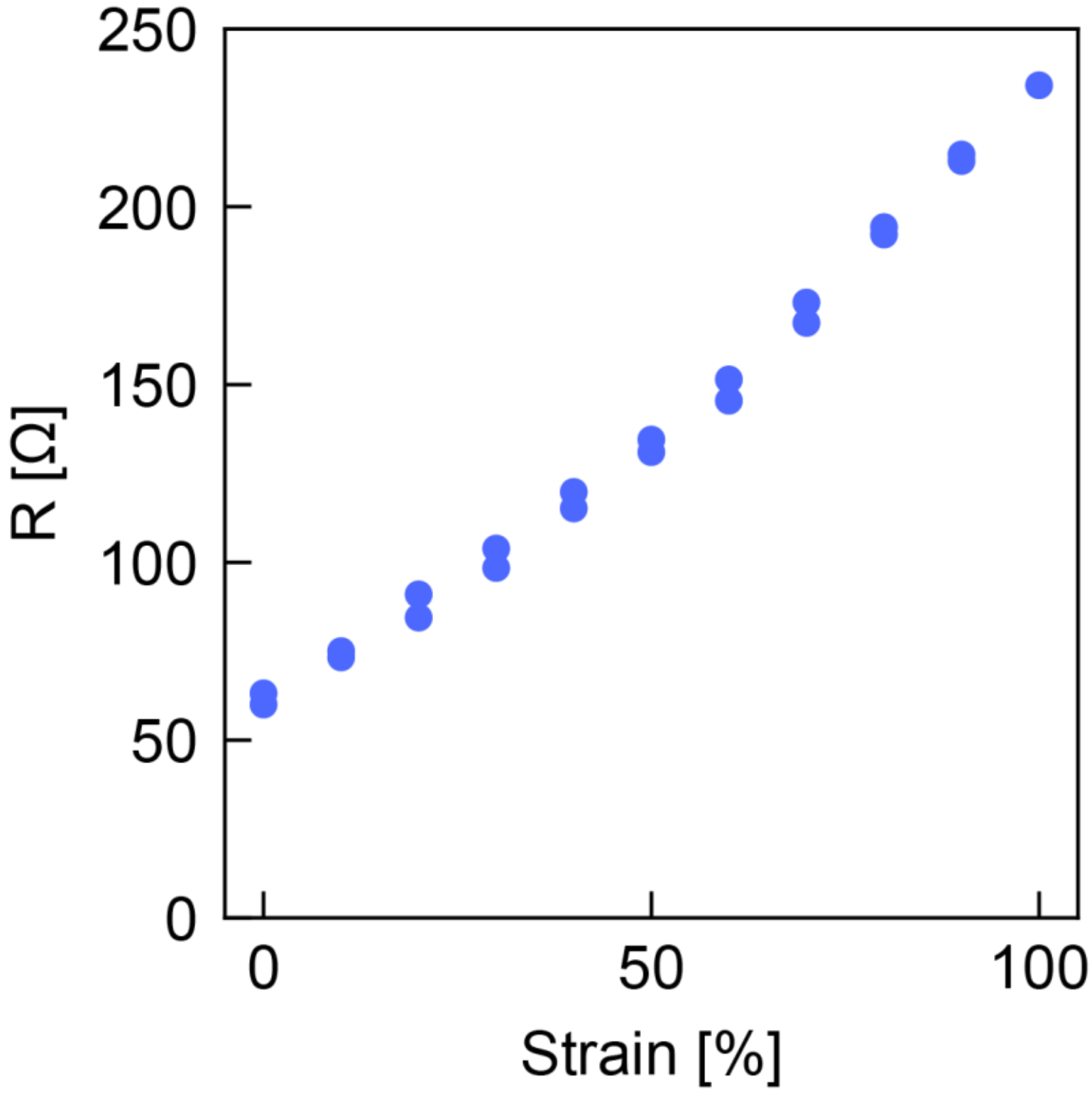

Figure S2. Strain–resistance relationship of liquid metal wirings. A linear, low-hysteresis change in resistance is observed over the range from 0 to 100% strain.

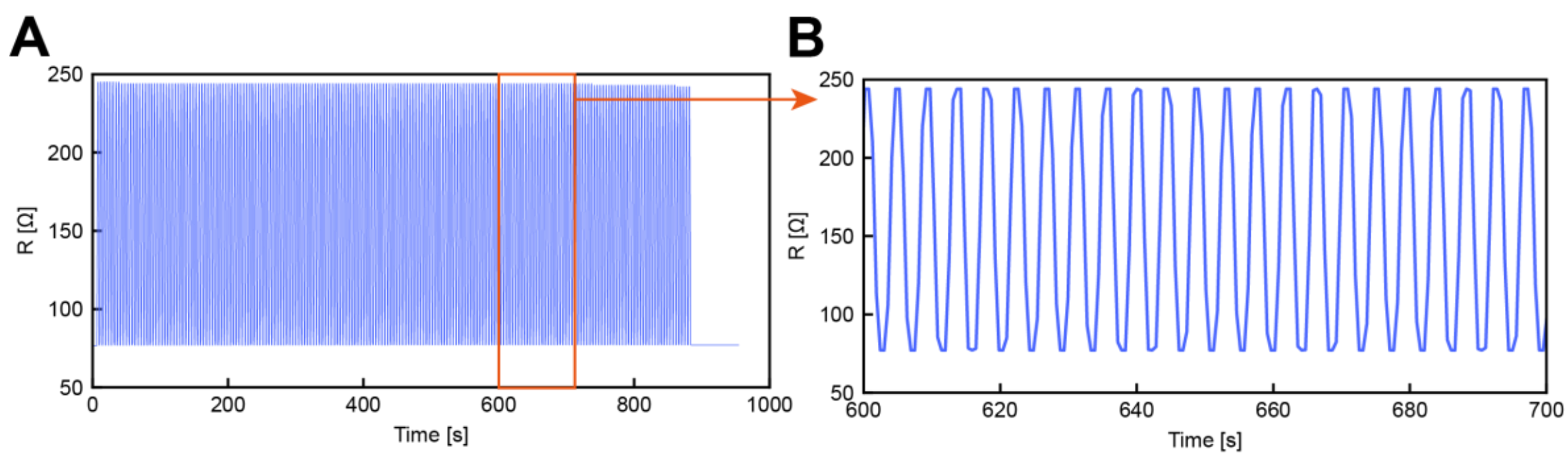

Figure S3. Electrical characteristics of liquid metal wirings during cyclic deformation. A. Strain–resistance relationship after 200 cycles of 100% strain. B. Magnified view.

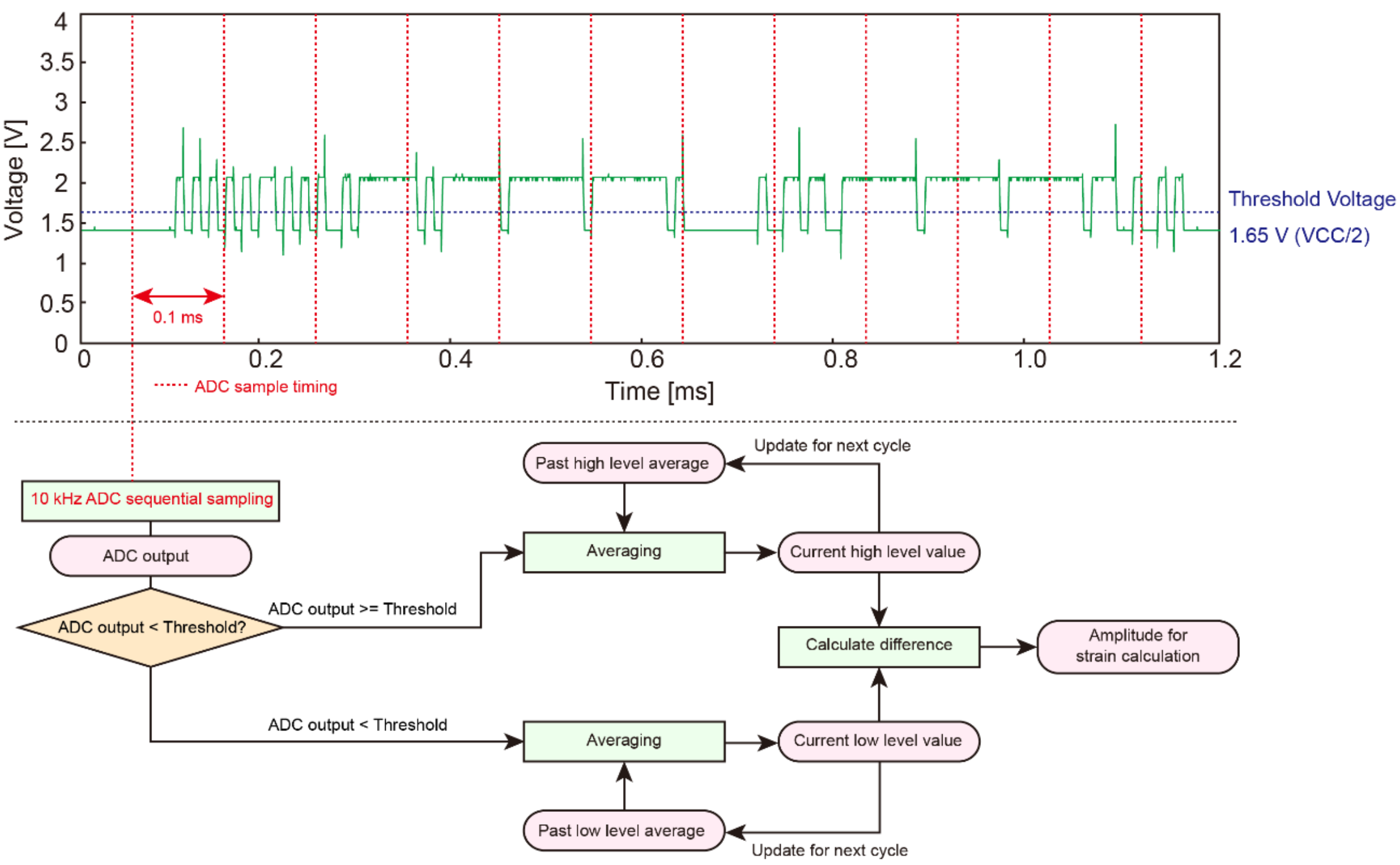

Figure S4. Procedure for reading the analog amplitude. Sampling is performed at regular intervals; using VCC/2 as the threshold, values above the threshold are treated as high and values below the threshold as low. Each value is averaged with its immediately preceding value, and the amplitude is finally output by subtraction.

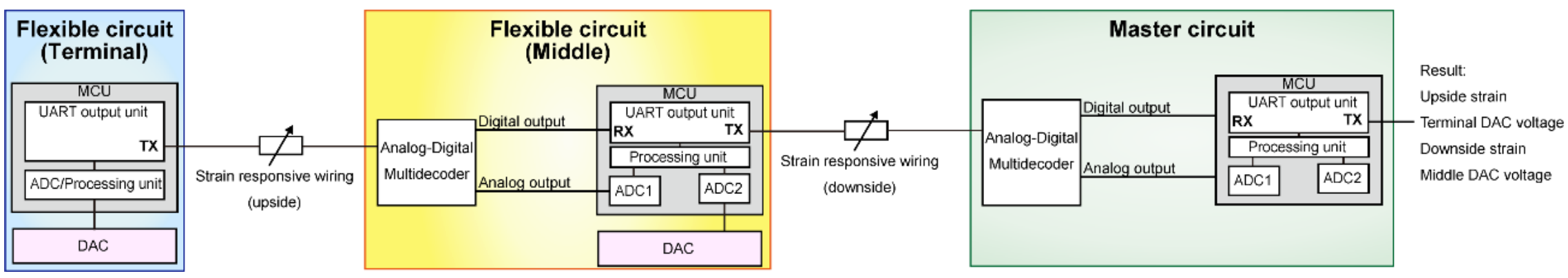

Figure S5. Explanation of the crosstalk measurement experiment using three-node serially connected devices (shown in Figure 2G). The intermediate unit reads the data it transmits, the data from the terminal unit, and the wiring strain between the terminal and intermediate units. This unit then re-encodes the data and transmits them to the master unit. By performing this digital re-conversion via the relay unit, the wiring required to directly connect the master and terminal units becomes unnecessary.

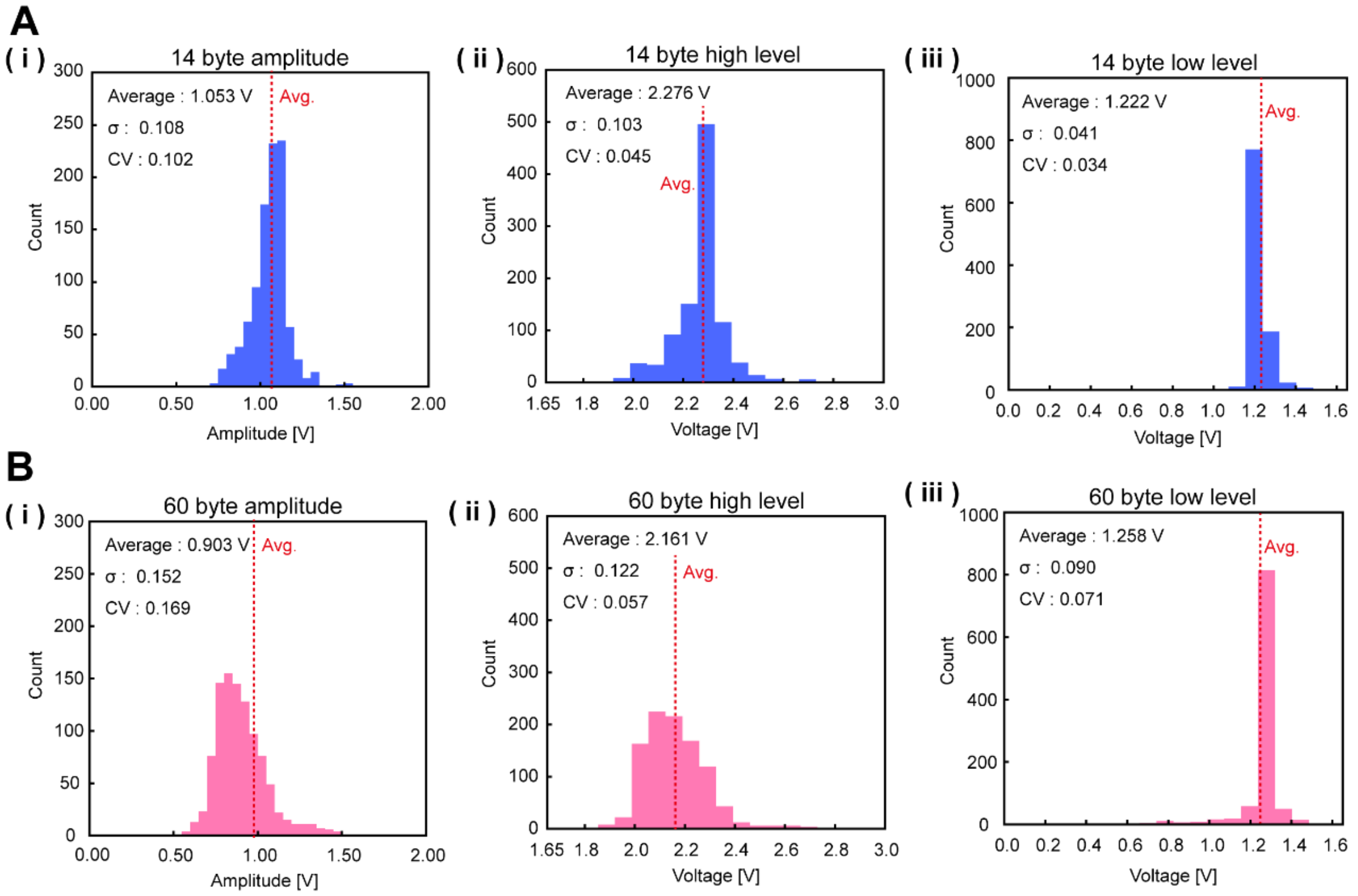

Figure S6. Distribution of the packet length, amplitude, high–level, and low–level voltages of digital data at 0% strain. CV: Coefficient of variation. A. 14-byte data. B. 60-byte data. (i) Amplitude. (ii) HIGH level. (iii) LOW level.

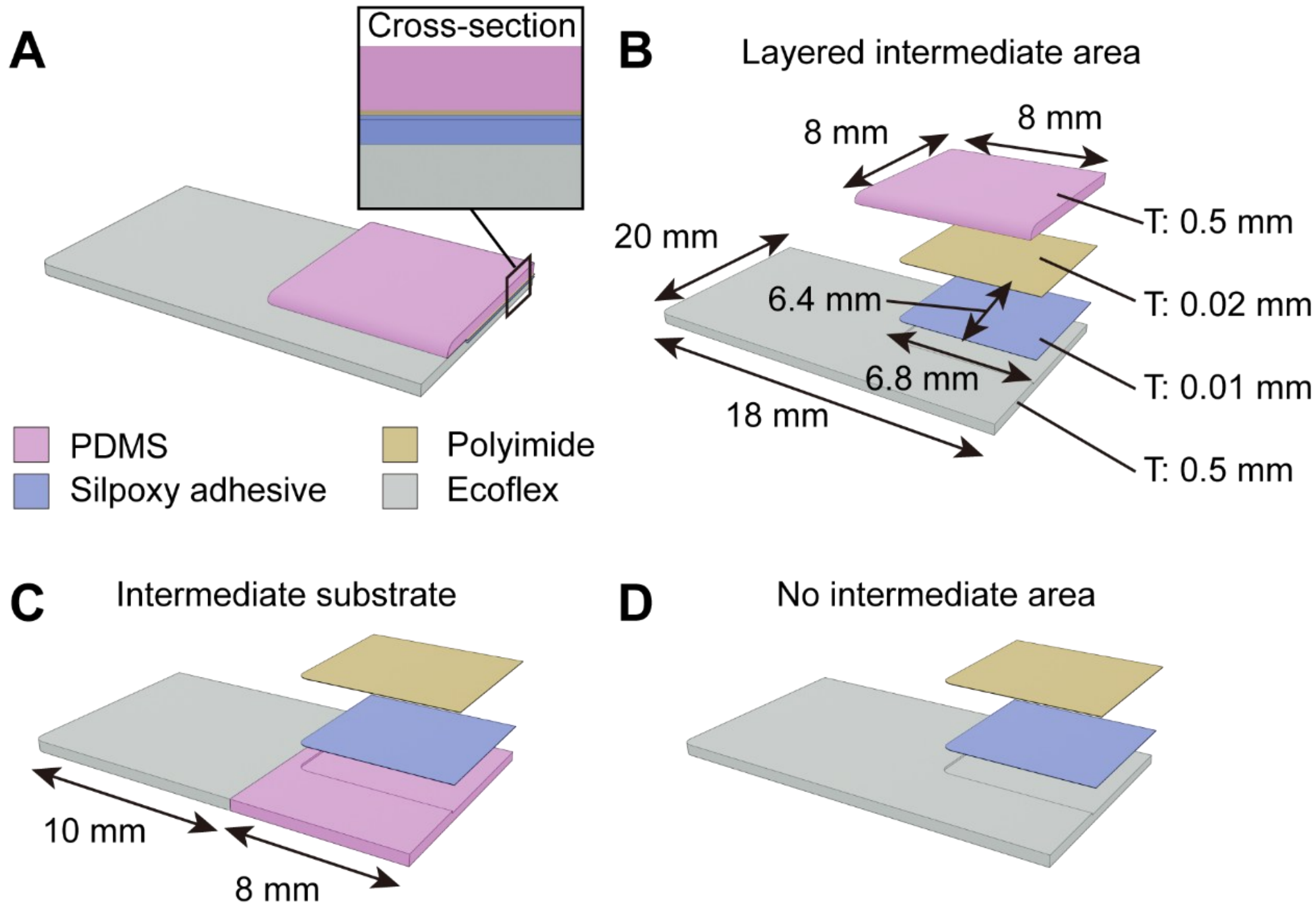

Figure S7. Geometry definition for the finite element analysis (FEA) simulation. The simulation was performed on a geometry one-quarter the size of the actual part, subject to left–right and top–bottom symmetry constraints. T: thickness. A. Overview. B. Layered structure. C. Substrate-embedded intermediate layer. D. Substrate without an intermediate layer.

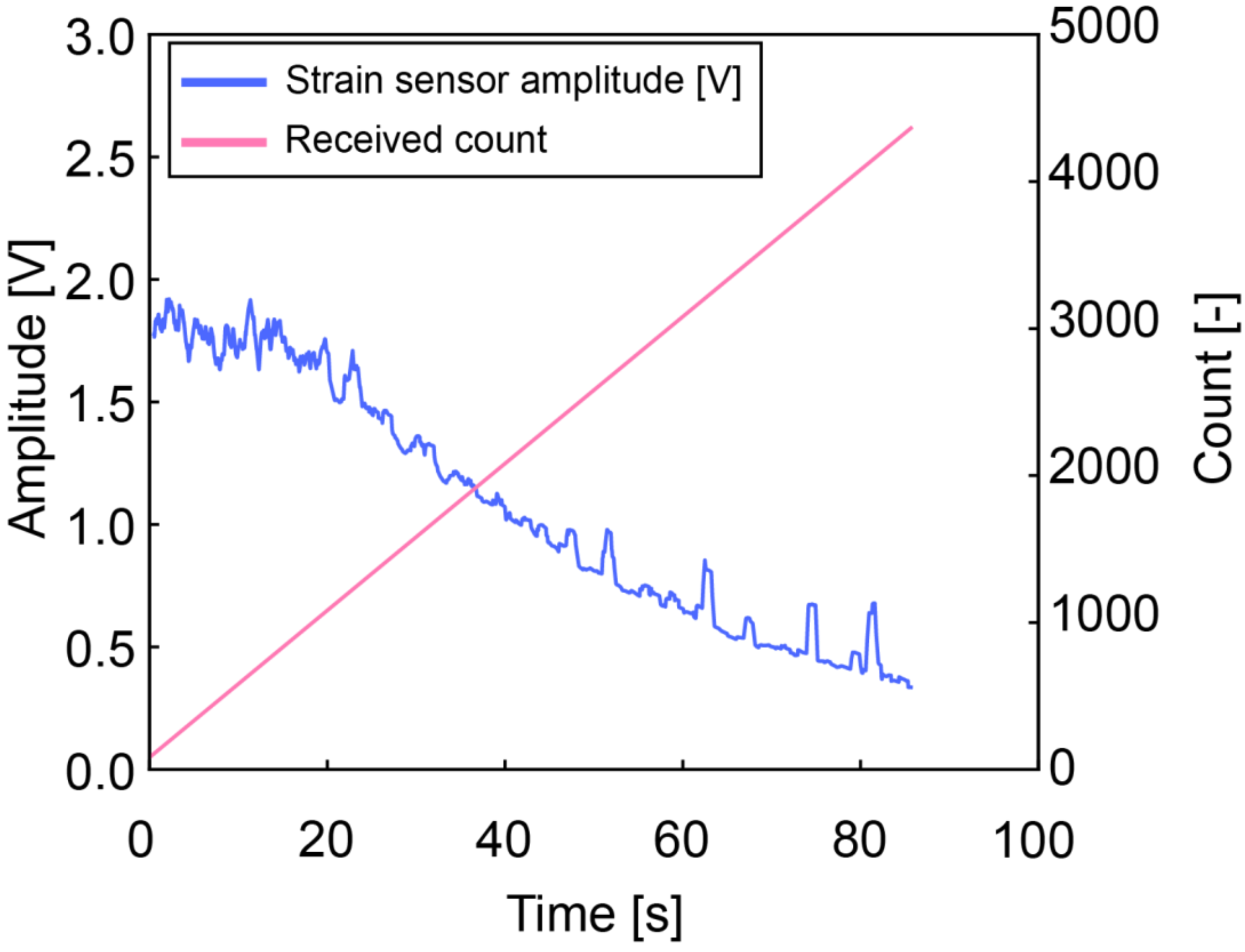

Figure S8. Log of the analog amplitude and digital communication during the communication-while-deformation test (shown in Figure 3G). The analog amplitude was smoothed using a 1-s moving average. The amplitude of the analog signal attenuated owing to deformation, whereas the digitally output count-up data were received without any loss.

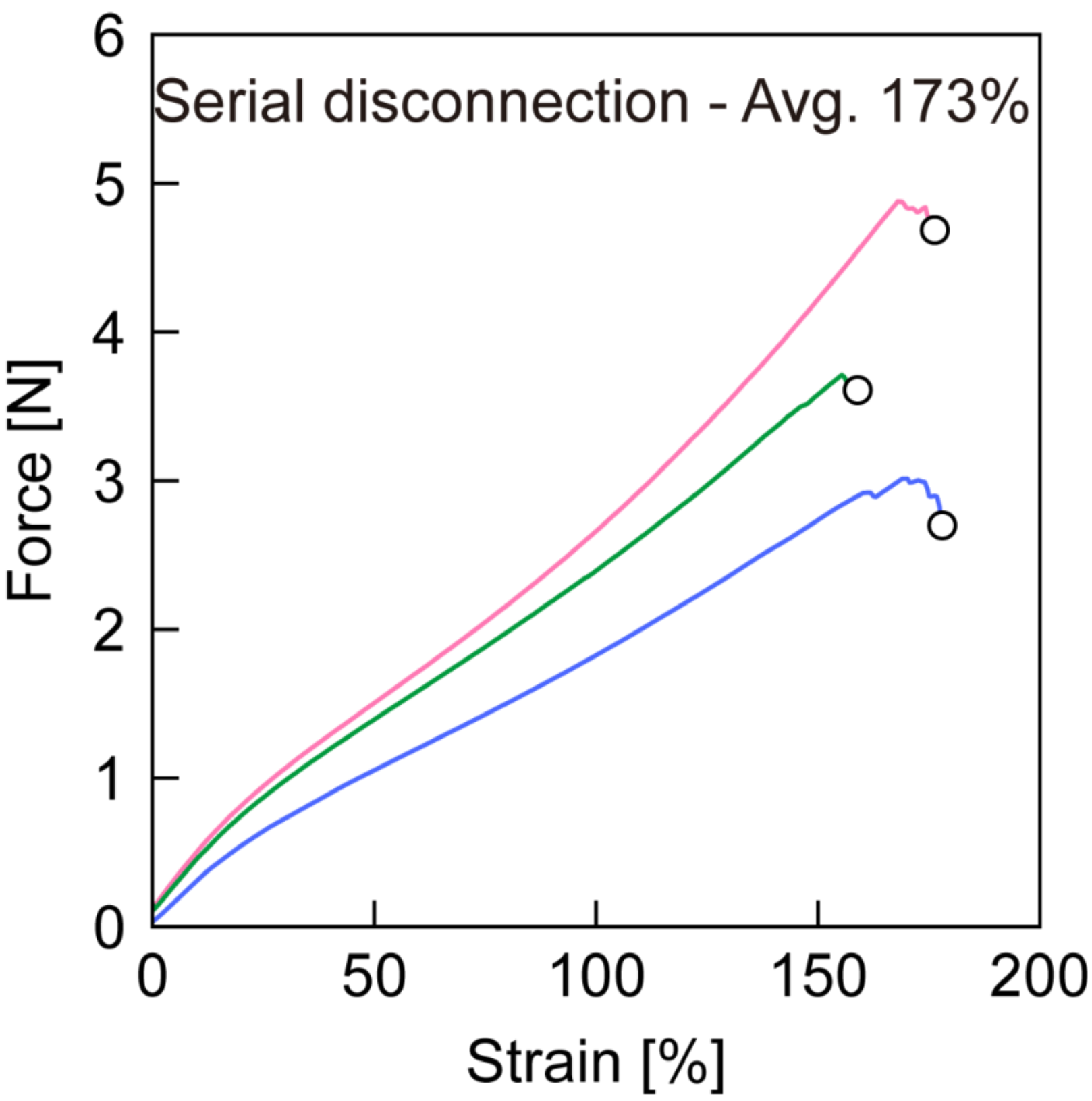

Figure S9. Tensile test result indicating communication failure in the devices without a protective layer.

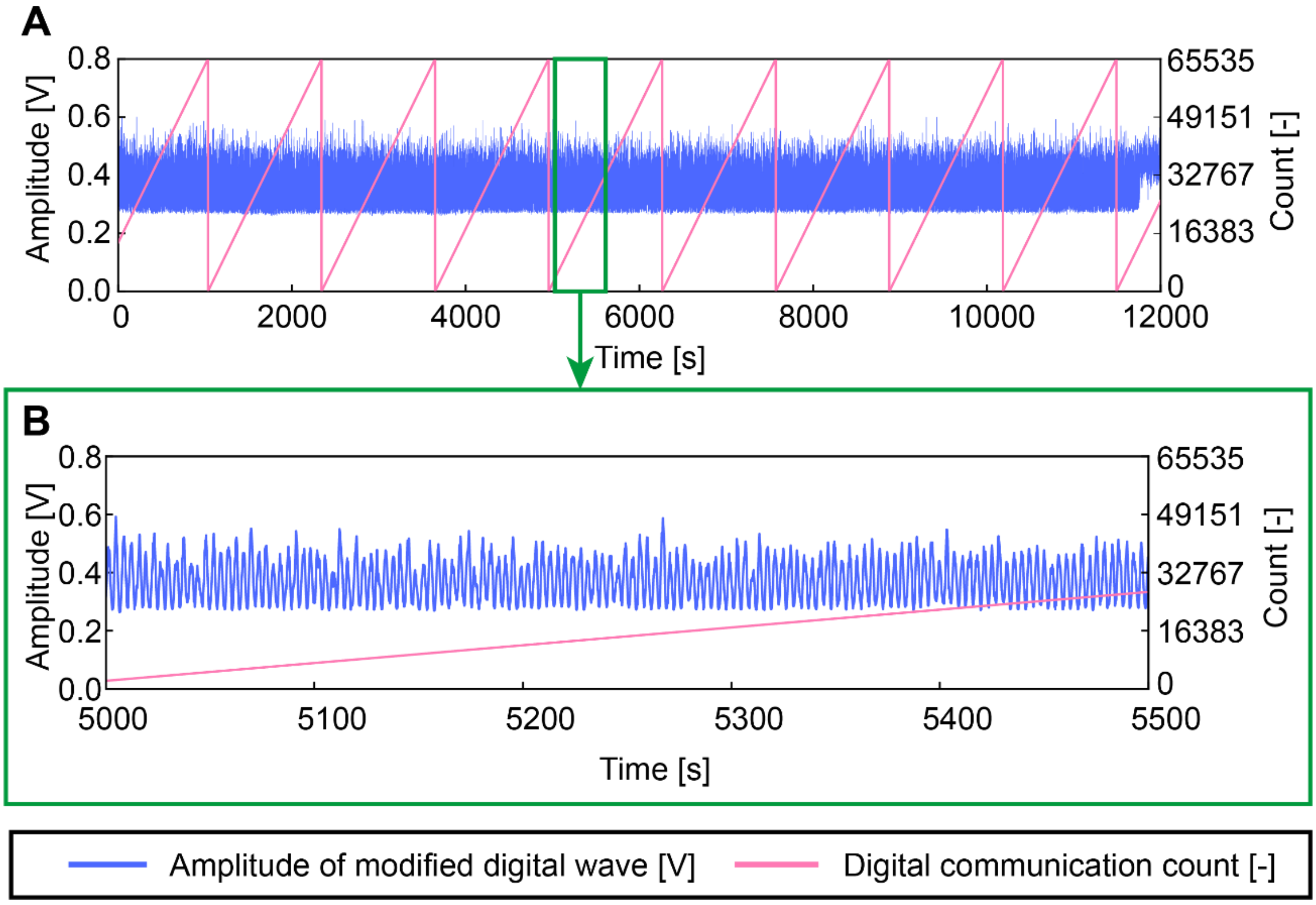

Figure S10. Log of the analog amplitude and digital communication during a repetitive strain test while communicating (shown in Figure 3H). The analog amplitude was smoothed using a 1-s moving average. The digitally output count-up data were received without any missing values and overflowed when the count reached the limit of a 16-bit integer.

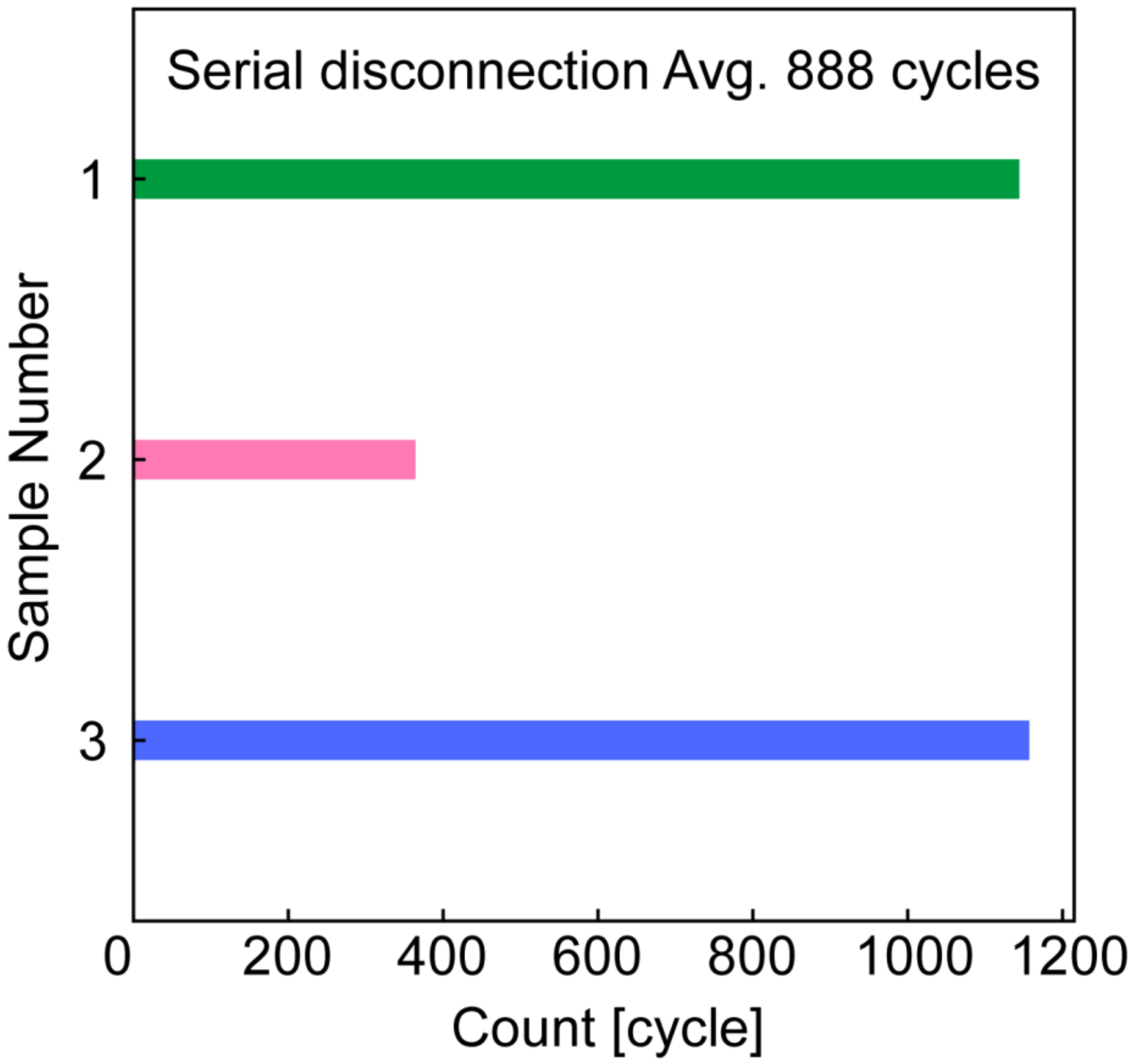

Figure S11. Number of cycles after which communication failure occurred in devices without a protective layer under 100% cyclic strain.

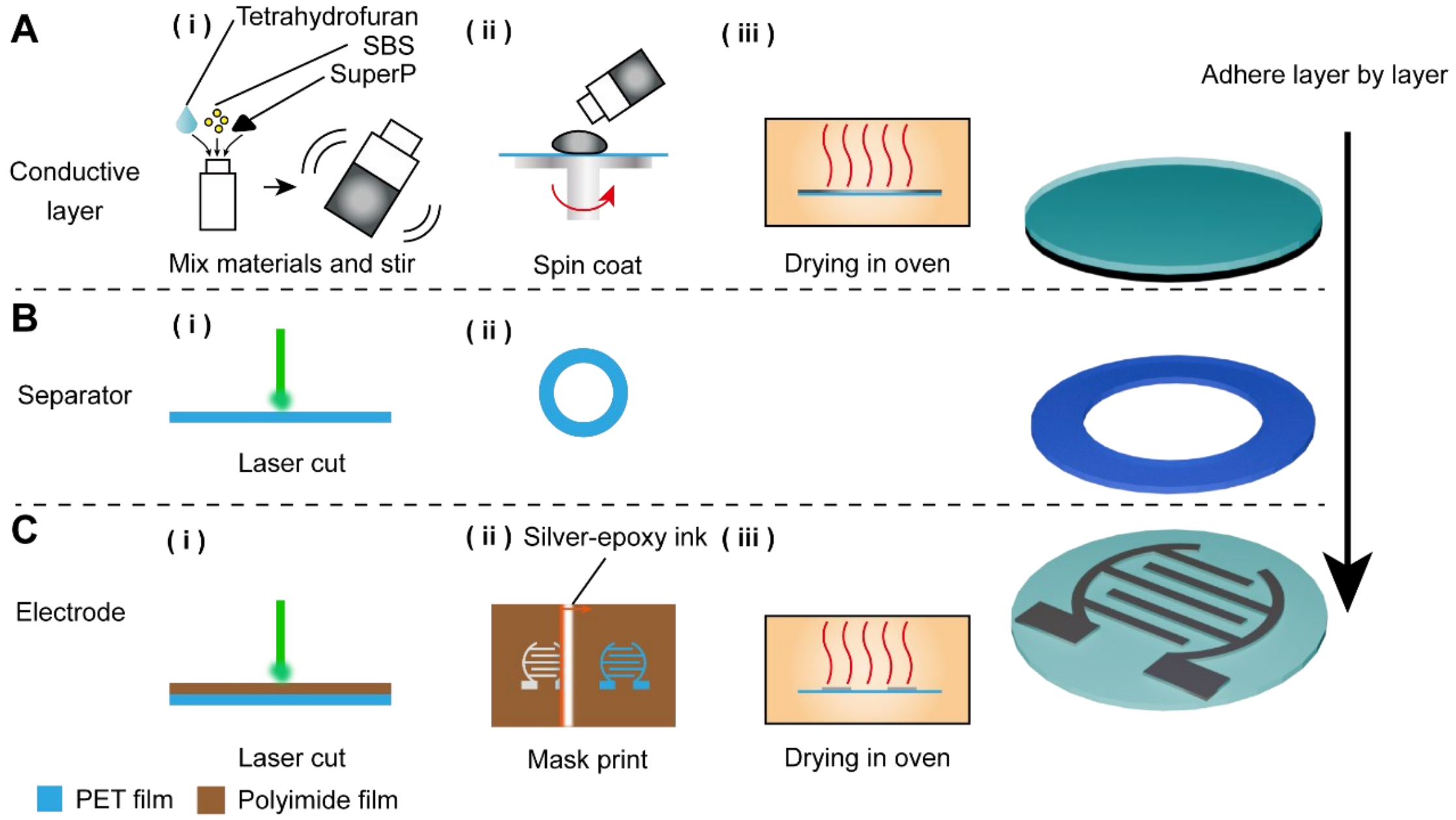

Figure S12. Fabrication process of the pressure sensor. The pressure sensor comprised a conductive composite (A), a separator (B), and an electrode (C).

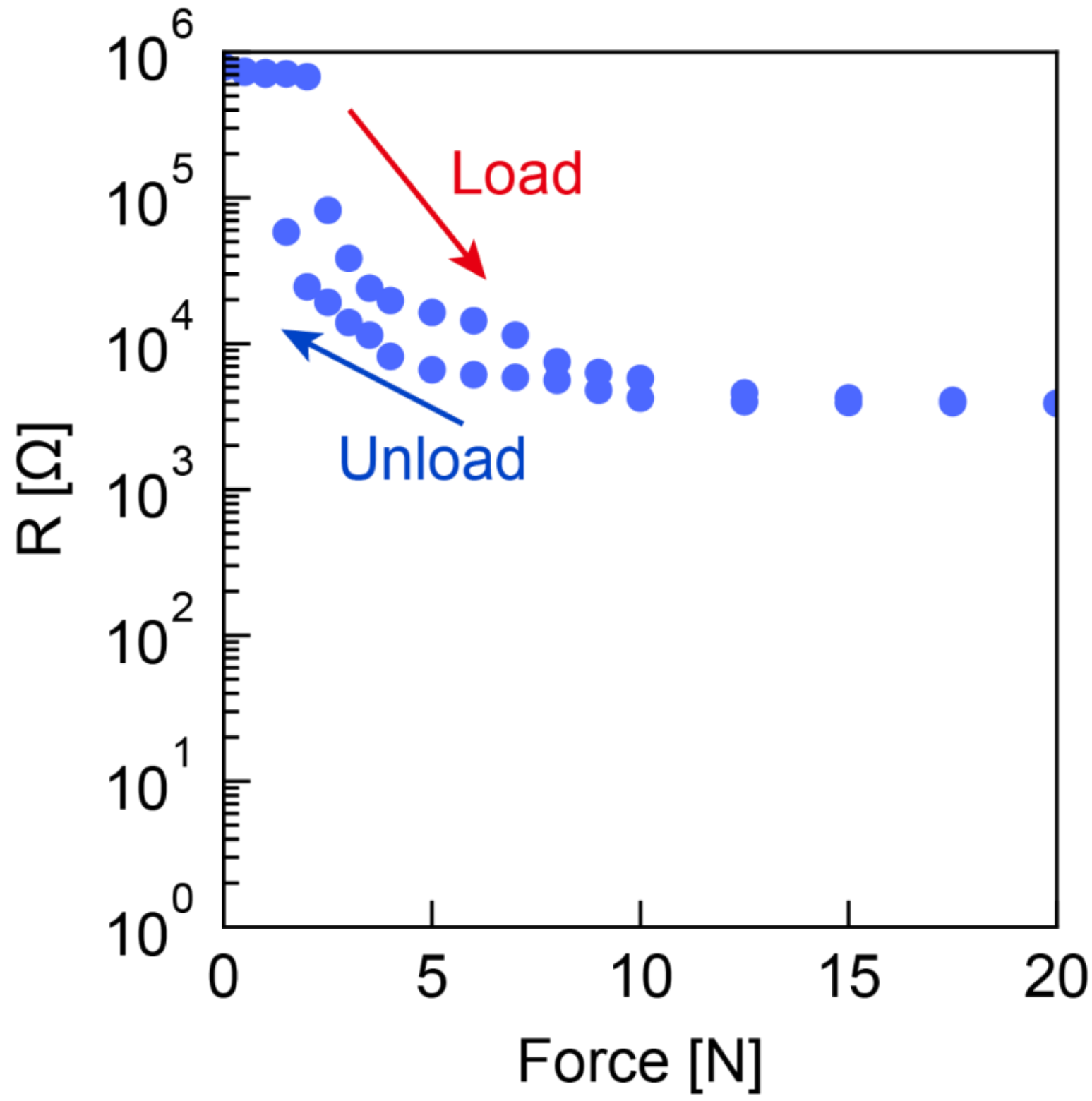

Figure S13. Force–resistance relationship of the fabricated pressure sensor.

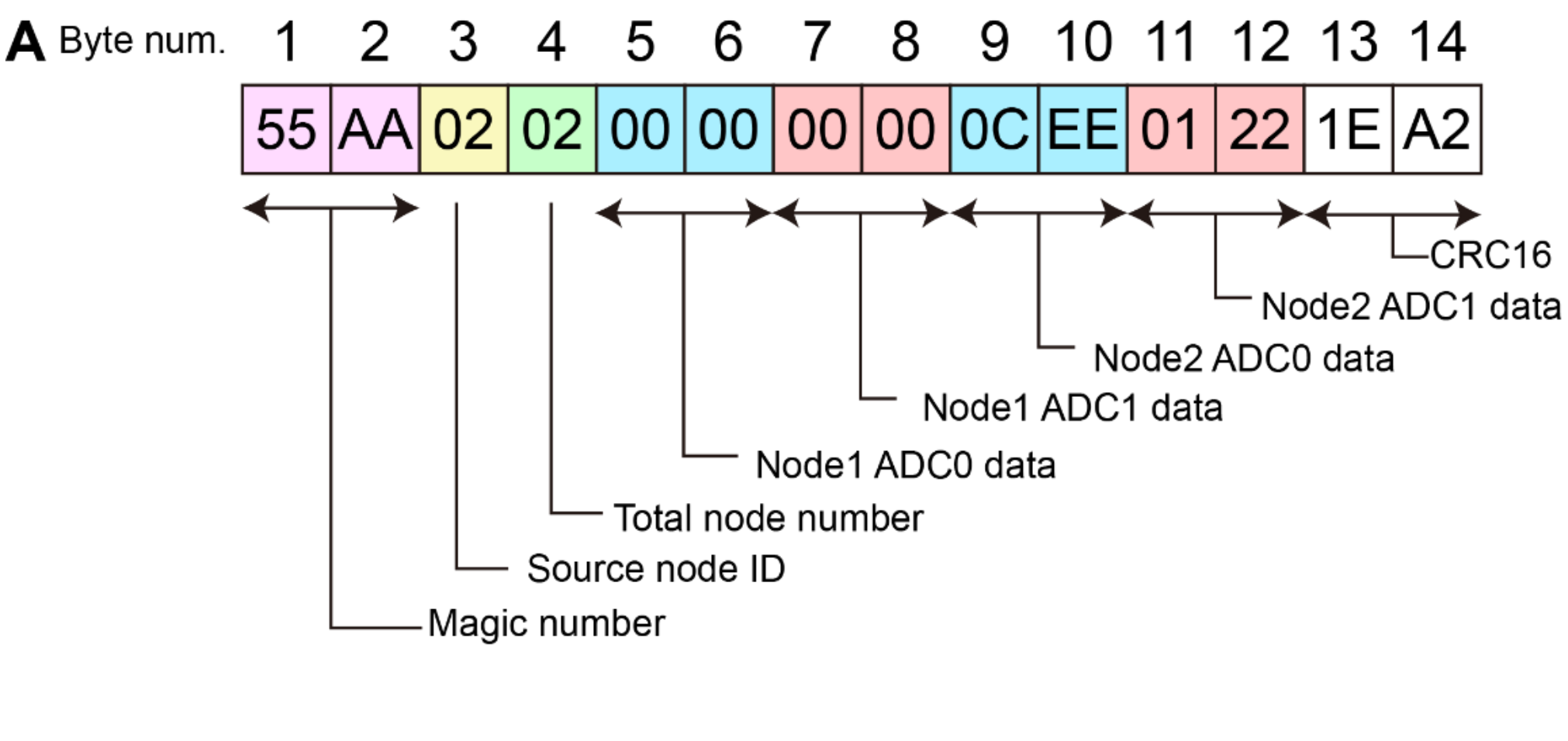

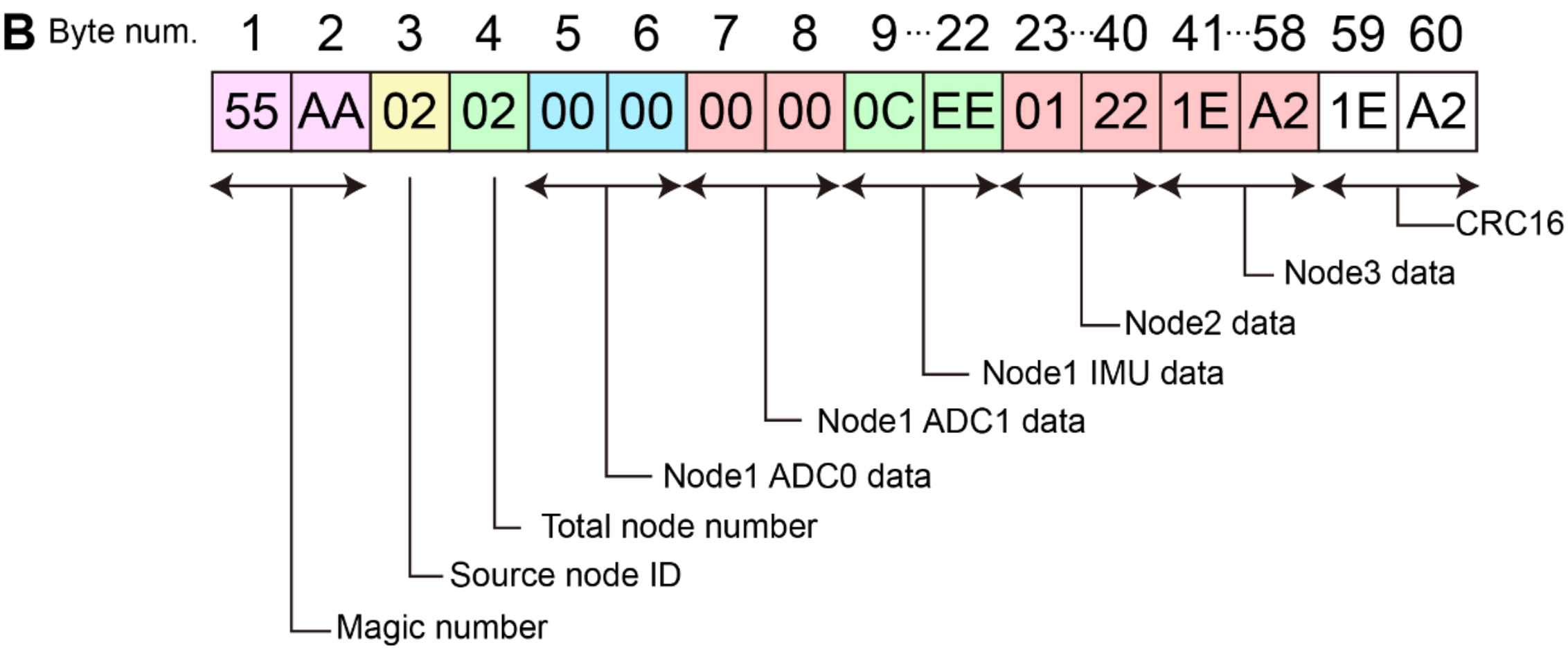

Figure S14. Packet structure for the multimodal communication architecture. A. Finger-pressure bending measurement. B. Self-shape mapping. A packet comprises a magic number (0x55AA) that indicates the start of the packet, the source node ID, the total number of nodes, and the data for each node and ends with a CRC16 value used to verify the overall integrity of the packet.

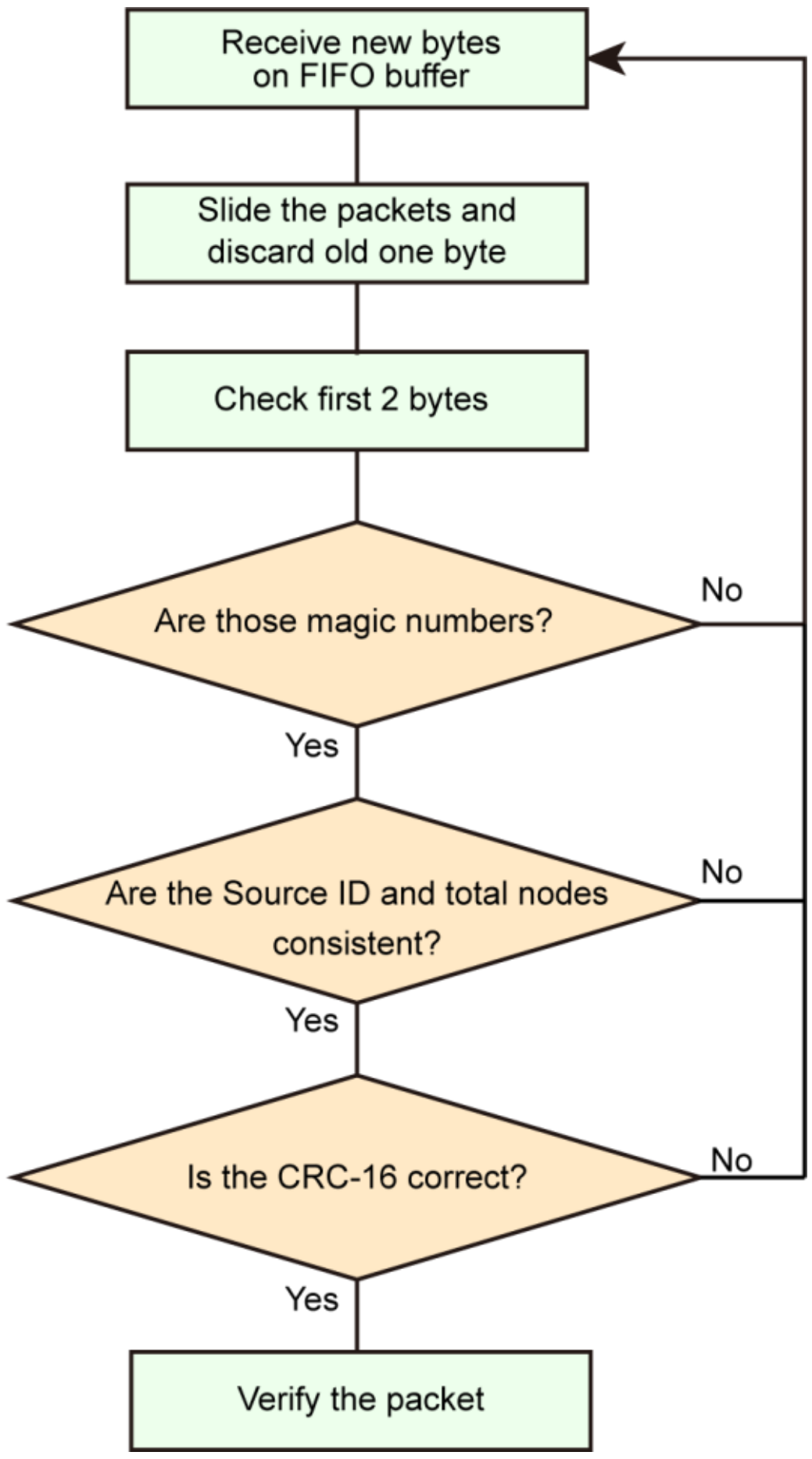

Figure S15. Flowchart for verifying digital communication reception. Data are received by shifting the buffer one byte at a time, and the packet integrity is verified by checking the magic number, source ID, total number of nodes, and CRC16.

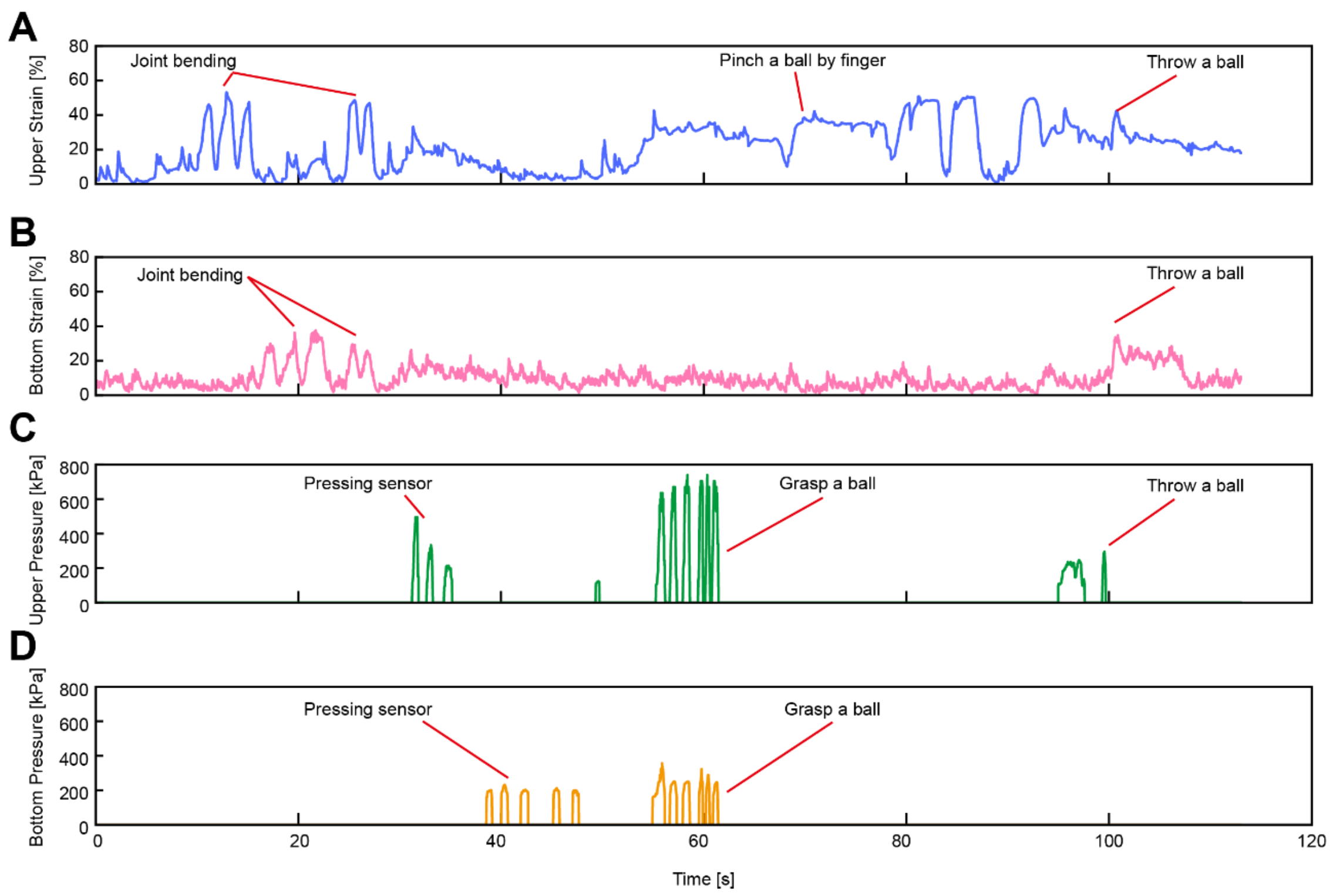

Figure S16. Log from the simultaneous measurement of finger pressure and bending demonstration (shown in Figure 4A–D, Supplemental Video 1). A. Proximal interphalangeal joint strain, B. Metacarpophalangeal joint strain, C. Fingertip pressure, and D. Middle-finger pressure.

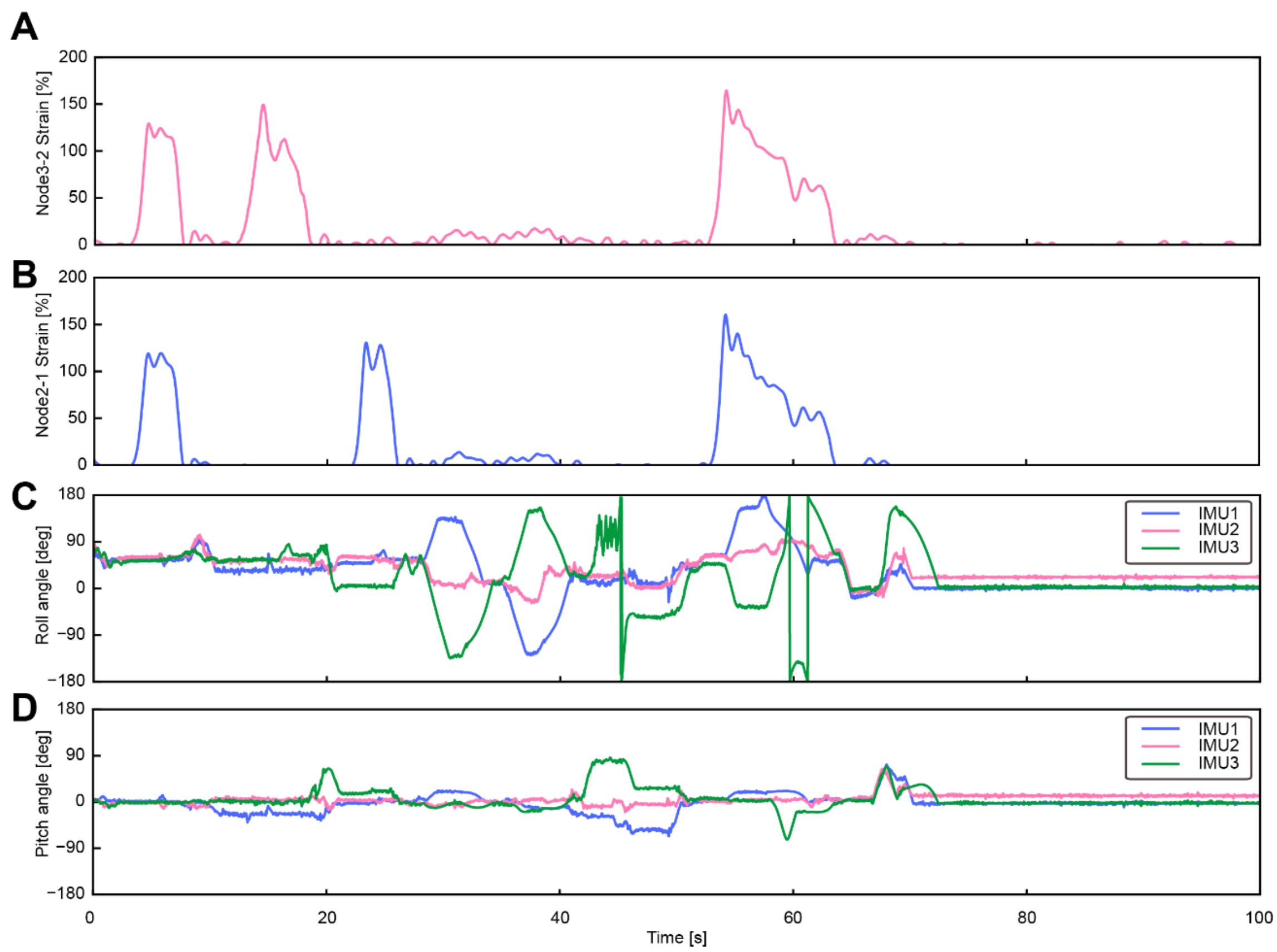

Figure S17. Log from the self-shape mapping (shown in Figure 4E and F, Supplemental Video 2). A. Strain between the terminal and intermediate nodes, B. Strain between the intermediate and master nodes, and C, D. Roll (C) and pitch angle of each node (D).

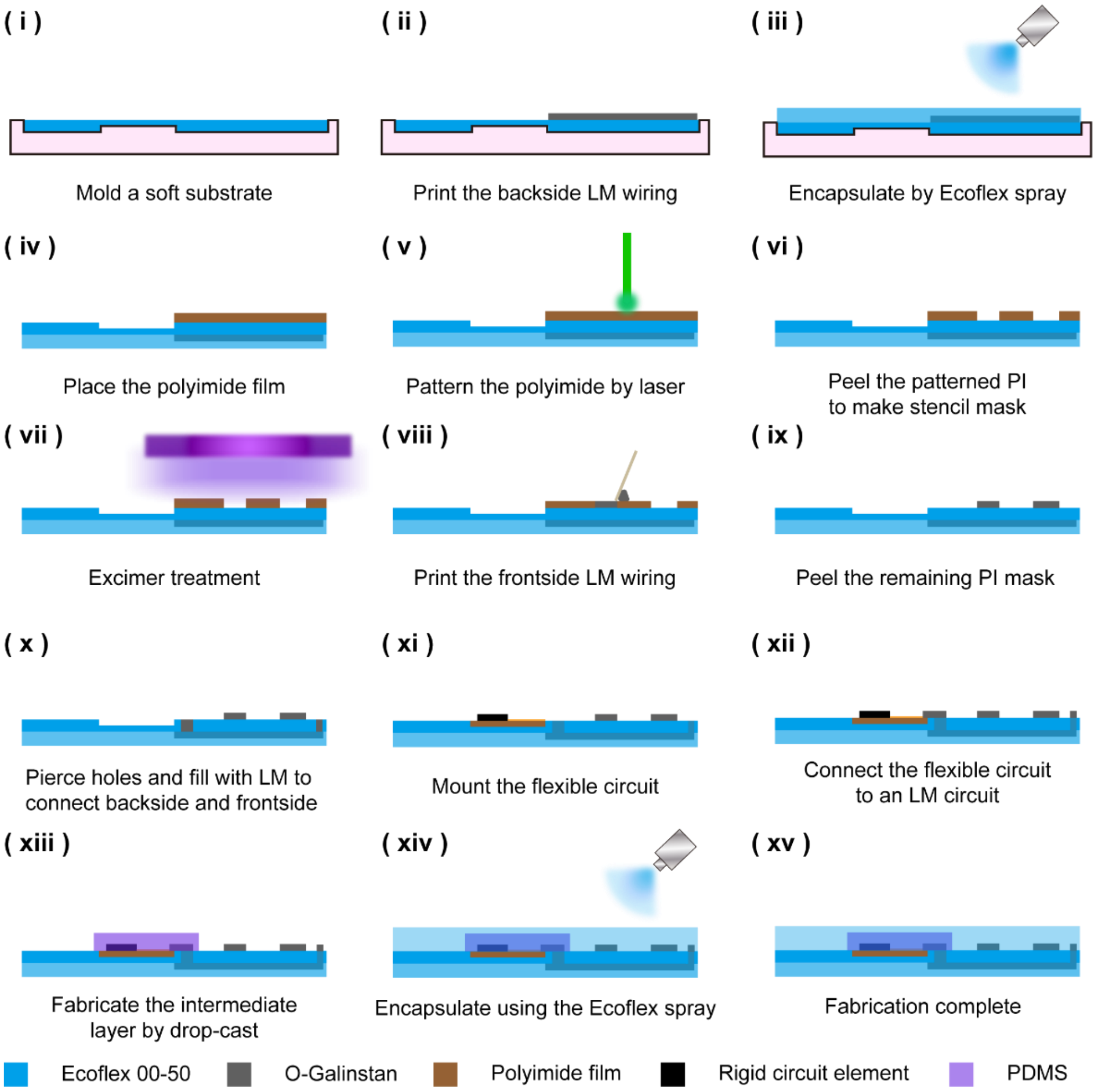

Figure S18. Device fabrication process. (i) Fabrication of a flexible substrate by molding. (ii) Printing of the liquid metal source wiring on the back side. (iii) Encapsulation of the backside wiring. (iv), (v), (vi) Fabrication of a stencil mask using a polyimide film. (vii) Excimer treatment to improve the liquid metal wettability. (viii) Stencil printing of the liquid metal. (ix) Removal of the mask. (x) Formation of through-holes in the encapsulation layer and connection of the top and bottom wiring. (xi) Mounting of the flexible printed circuit board (PCB). (xii) Connection of the flexible PCB to the liquid metal wiring. (xiii) Fabrication of the intermediate layer. (xiv) Encapsulation of the top surface. (xv) Completion of processing.